\documentclass[fleqn,10pt]{wlscirep}
\usepackage[utf8]{inputenc}
\usepackage[T1]{fontenc}
\usepackage{lineno}
\usepackage{multirow}
\usepackage{geometry}
\usepackage{pdflscape}
\usepackage{caption}
\usepackage{siunitx}
\usepackage{subcaption}
\usepackage{tabularx}
\usepackage{xr}

\makeatletter
\newcommand*{\addFileDependency}[1]{% argument=file name and extension
  \typeout{(#1)}
  \@addtofilelist{#1}
  \IfFileExists{#1}{}{\typeout{No file #1.}}
}
\makeatother

\newcommand*{\myexternaldocument}[1]{%
    \externaldocument{#1}%
    \addFileDependency{#1.tex}%
    \addFileDependency{#1.aux}%
}

\newcolumntype{C}{>{\centering\arraybackslash}X} 

\myexternaldocument{supplementary}

\title{Impact of demand growth on decarbonizing India's electricity sector and the role for energy storage}

\author[1]{Marc Barbar}
\author[1,*]{Dharik S. Mallapragada}
\author[1]{Robert Stoner}
\affil[1]{Massachusetts Institute of Technology, Cambridge, MA, 02139, United States}

\affil[*]{dharik@mit.edu}

\keywords{Energy storage, space cooling, decarbonization, emerging markets}

\begin{abstract}
Global energy sector decarbonization efforts are contingent on technology choices for energy production and end-use in emerging markets such as India, where air conditioning is expected to be a major driver for electricity demand growth. Here, we use an integrated demand-supply framework to quantify the impacts of demand growth and temporal patterns on long-term electricity system evolution. Under projected renewables and Li-ion storage cost declines, our supply-demand modeling points to renewables contributing substantially (46-67\%) to meet annual electricity demand in India by 2030. However, without appropriate policy measures to phase out existing coal generation, even such rapid adoption of renewable energy coupled with one or more technological levers such as low-cost energy storage and demand-side measures such as setting aggressive AC efficiency standards and deploying distribution level storage, are insufficient to reduce annual CO$_2$ emissions in 2050 vs. 2020 because of the relatively higher growth rate of projected electricity demand over this period. 

\end{abstract}
\begin{document}

\flushbottom
\maketitle
\thispagestyle{empty}

\section*{Introduction}
Electricity generation in emerging markets and developing economies (EMDE), such as India, Indonesia, Brazil, South Africa and Nigeria \cite{Rao2019} over the next few decades will significantly impact global greenhouse gas (GHG) emissions as access and economic development increase electricity demand in these regions\cite{su10051440, McNeil2013}. Notably, because many of these countries are located in hot climate zones, they are expected to see growing energy demand for space cooling, which in many cases, is likely to be exacerbated by climate change impacts \cite{Barbar2021, Waite2017, Colelli2020}. For EMDE as a whole, more investigative planning of energy infrastructure from both the demand \cite{Zhou2018, Muratori2018} and supply \cite{Debnath2018} perspective is warranted to ensure efficient use of limited capital and alignment with global mid-century climate mitigation goals. Of these countries, India stands out since it already ranked 3$^{rd}$ in terms of CO$_2$ emissions in 2018 \cite{IEAEnegyAtlas}, owing to its large population and reliance on coal for primary energy (44\% of primary energy demand in 2019\cite{IEAWEO2020}), and in particular for electricity generation (72\% of supply\cite{IEAWEO2020}). Yet, these national statistics belies the lower annual per-capita primary energy consumption in India (23 million btu (MMBtu)) compared to other high-income countries like United States (310 MMBtu) and Germany (165 MMBtu) in 2018 \cite{IEAWEO2020}. By one estimate, growth-driven energy consumption could result in India's final energy use in 2040 being 81\% greater than in 2019 \cite{IEAWEO2020}, with demand for electricity growing much faster in this scenario than other forms of energy, at 161\% \cite{IEAWEO2020}. While decadal electricity demand growth projections for developed countries such as the U.S. is driven primarily by the electrification of transportation \cite{NRELEFS2018}, in India and many other EMDE countries with hot climates, the building sector is projected to dominate electricity demand growth over this period, primarily due to the widespread adoption of air conditioning (AC) systems. As compared to other new sources of demand (e.g., EVs), the relative inflexibility and timing of AC use means that it will not only increase aggregate demand but also change the temporal load shape and impact peak consumption. For example, a recent study estimates that space cooling could contribute as much as 45\% of peak electricity demand in India by 2050 compared to 10\% in 2016 \cite{IEAcooling18}. In the short-term, increases in peak demand, which tends to occur after sunset\cite{NREL2019}, will likely be met with relatively high emissions intensity coal-based electricity generation \cite{Mallapragada2018}, owing to its dominant share of supply today. Assessing pathways for grid decarbonization in the Indian context and other similar regions, therefore requires a granular study of the temporal patterns of AC demand in conjunction with dynamics of electricity generation. 

Recently, several studies have analyzed the operation and long-term evolution of India's bulk power system between 2030 and 2050 at different levels of granularity in representing grid operations, existing generation, evolution of demand profile, and investments in new generation, storage and transmission \cite{Deshmukhe2008128118, Lu2020, NREL2019, NREL2020, TERI2020, Rudnick2019, He2016, Kittner2017, Craig_2018}. Some studies model grid operations for various generation capacity scenarios in 2030 to quantify the operational feasibility of different levels of variable renewable energy (VRE) penetration and the flexibility provided by coal and hydro generation as well as new battery storage to integrate VRE generation \cite{Deshmukhe2008128118, TERI2020, NREL2019}. Other studies model the long-term evolution of India's electrical grid (to 2050) \cite{Malik2020, NREL2020, Rudnick2019, Lu2020} subject to approximations regarding spatial and temporal variations in demand and VRE resource characterization and its impact on capacity investment. Notably, the temporal resolution of grid operation in these studies range from a few time periods (<50) per year at a state-level spatial resolution \cite{NREL2020, Malik2020} to hourly operations at the regional level\cite{Rudnick2019, Lu2020}. In addition, some studies use simplified modeling based on a single resource profile per region \cite{Rudnick2019} while others use detailed representations of VRE resource availability that may include land availability \cite{Lu2020} and transmission interconnection costs \cite{NREL2020}. Some studies also model investment over multiple periods and the temporal evolution of the power system from 2020 to 2050 \cite{Malik2020, NREL2020}. A key finding across many of these studies is the cost-effectiveness of VRE generation deployment in the future Indian electricity system (>50\% as much as 80\% of annual generation\cite{Lu2020}). However, none of these investment planning studies account for the structural changes in the electricity consumption profile for electricity use over time, resulting from factors such as AC adoption, and to a lesser extent EV adoption. Therefore, these studies present an incomplete picture of long-term evolution of the power system in India.

Our contribution improves upon prior work by developing a holistic framework for assessing the impact of supply \textit{and} demand-side drivers on the long-term evolution of the power sector in India and other EMDE regions. This supply-demand interaction is addressed by combining bottom-up demand forecasting with high temporal resolution capacity expansion modeling (CEM) that uses high spatial resolution VRE resource availability and detailed representation of hourly grid operations\cite{Jenkins2017}. The bottom-up demand forecasting model, documented in detail elsewhere\cite{Barbar2021}, captures the growth of business-as-usual components as well as new components, namely AC and EVs, in estimating electricity demand at the state-level in future years at an hourly resolution. This granularity enables us to explore the system impact of demand-side interventions, such as improved AC efficiency standards, alternative EV charging schedules as well as the potential impact of distribution level energy storage (DLS) deployment to manage congestion in the local distribution system. The resulting regional demand profiles are subsequently used as inputs to a multi-period power system CEM that considers grid operations at an hourly resolution, to evaluate the least-cost trajectory of power system investment and operation in India from 2020 to 2050. This framework, (see Supplementary Fig. \ref{fig:schematic}), is used to address the following questions in this study: a) How do various demand-side drivers (AC, EV load growth) impact the evolution of India's power system in terms of generation capacity mix and CO$_2$ emissions when factoring interactions with long-term supply-side factors such as natural gas (NG) prices, VRE availability and energy storage  capital costs? and b) How does AC demand growth impact the need for energy storage, at distribution and transmission levels, and both existing and new coal generation under various technology and policy scenarios?

\section*{Methods}\label{sec:methods}

\subsection*{Demand-side scenario model}\label{method:demand}

The alternative electricity demand scenarios evaluated here are developed using a previously documented open-source model \cite{Barbar2021} that uses separate approaches to estimate future electricity demand for existing end-uses ("business-as-usual" model) as well as demands from emerging end-uses such as ACs and EVs ("technology model"). Electricity demand from existing end-uses is estimated for future periods using a regression model that is trained on historical regional electricity demand available for 2012-2019\cite{posoco} at the daily resolution and hourly demand for 2015 \cite{Rudnick2019}. In addition, this model incorporates weather data at daily resolution and GDP forecasts at monthly resolution to incorporate seasonal trends and long-term growth respectively. 

The technology model enables a bottom-up approach to estimate demand from new loads, which in this study relates to space cooling in residential and commercial buildings as well as EV charging \cite{Barbar2021}. The model relies on AC sales data projection as well as types of units being sold to meet the expected space cooling demand. Two AC scenarios were considered: a baseline scenario with electricity sales projections based on currently available AC units and a high efficiency scenario that assumes preferential adoption of efficient AC units as defined by a recent study \cite{IEAcooling18}, which considers a scenario where the global average Seasonal Energy Efficiency Ratio (SEER) rating of ACs reaches 8.5 by 2050 \footnote{although high AC efficiency scenario is defined here based on improvements in SEER ratings in the Indian context, it can also be viewed as the outcome of other building sector interventions like passive cooling that reduce overall electricity demand to achieve a similar level of thermal comfort as in the baseline scenario}. As of 2018, by comparison, the sales-weighted average SEER for ACs in India was 3 and the global average was 4 \cite{IEAcooling18}. \footnote{AC efficiency, as reflected in SEER ratings, may differ greatly between the United States and India due to the types of AC units installed. While heating, ventilation, and air conditioning (HVAC) systems use efficient cooling methods such as variable refrigeration in the U.S. context, their system cost is high for the Indian market where less efficient split units are expected to be installed}. Residential and commercial AC demand growth was estimated at the state level and then aggregated to the regional level \cite{Barbar2021} to be input to the supply-side optimization model. For EVs, the technology model uses vehicle sales data and government goals for EV sales targets in future years \cite{NITI2018} to estimate EV charging demand. Additionally, hourly projections of EV charging demand at the regional level were derived for each decade after applying a 1D-convolution to survey data related to typical charging patterns in an EMDE settings \cite{Barbar2021}. As compared to AC demand, electricity demand from EV charging is projected to be relatively modest, both in terms of annual consumption and in terms of contribution to peak demand, as seen in Table \ref{table:demand}. The reference electricity demand projection for our analysis is estimated assuming stable GDP growth, baseline AC efficiency, evening EV charging scheme. Evening EV charging is predominant in other EMDEs such as Mexico \cite{berkeley} and therefore is chosen as the schedule for the reference case.

\subsection*{Supply-side optimization model}\label{method:model}
We use a multi-period version of the power system CEM \cite{He2016, Craig_2018, Carvallo2017}, GenX \cite{Jenkins2017} to evaluate the least-cost investment and operation of the Indian power system under alternate technology, demand and policy scenarios. The GenX model is open-source \cite{gitgenx}.
%The GenX model has been used previously as a single period CEM  to study questions such as the value of energy storage integration \cite{Mallapragada2020}, the role for firm low-carbon resources like nuclear and thermal power generation with carbon capture \cite{Sepulveda2018} and the role and value of long-duration energy storage systems under deeply decarbonized grid scenarios \cite{Sepulveda2021}. 
For this study, GenX is configured as a multi-period investment planning model with four investment periods (2020, 2030, 2040, 2050) and hourly representation of grid operations. We solve the resulting linear programming (LP) model using a dual dynamic programming (DDP) algorithm that makes this problem computationally tractable by decomposing the problem into individual sub-problems per investment period and subsequently uses information from solution of the model in future investment periods ("Forward Pass") to adjust investment decisions in previous periods ("Backward Pass") \cite{Lara2018}. For each investment period, the model include the following grid operations constraints: a) flexibility limits of thermal power plant operations via linearized unit commitment constraints \cite{Palmintier2013, Poncelet2020}, b) supply-demand balance at each hourly time step and each zone, with power flow associated with linear losses and transfer capacity limits between zones, c) modeling hydro power plant operation to adhere to available information on inflows and reservoir capacity \cite{Rudnick2019} and d) modeling other storage resources with inter-temporal storage balance constraints as well as capacity constraints on maximum rate of charging and discharging.
%For electrochemical battery storage (metal-air) and long-duration storage like power to hydrogen (H$_2$) to power, we allow for power and energy capacity to be independently sized to maximize their value to the power system while adhering to the storage design constraints, if any. 
These operational constraints are modeled over 20 representative weeks of grid operation, that are selected from an single year of load data based on 2015 weather patterns, VRE and hydro resource profiles (more details discussed later on) via k-means based clustering \cite{Mallapragada2020, Mallapragada2018a}. The operations over the 20 representative weeks are scaled up to estimate annual operation cost and other operational metrics of interest such as VRE curtailment and CO$_2$ emissions. The choice of 20 representative weeks was made to balance accuracy of capturing intra-annual variability in load and VRE, hydro resource availability as well as computational run times for the multi-period CEM (see Supplementary Fig. \ref{fig:15_SI}). Additionally, Supplementary Fig. \ref{fig:20_SI} highlights the relative error in capacity outcomes compared to the model with 20 representative weeks as the baseline.

We represent the Indian grid using five separate balancing regions (North, West, South, East and Northeast) defined by the grid operator \cite{CEA2018}, with region-specific load profiles developed for each investment period based on the above-mentioned demand-forecasting model \cite{Barbar2021} - example outcomes are shown in Supplementary Fig. \ref{fig:18_SI}. The power flows between these regions are modeled based on a simplified network representation that enforces power exchange limits between the regions (see Supplementary Table \ref{table:network}). For 2020, these power limits are derived from the grid operator \cite{Rudnick2019, CEA2018}. These limits may be expanded with additional transmission investment in future periods.

\subsection*{Resource cost and performance assumptions}\label{method:assum}
GenX models operations over four periods, 2020, 2030, 2040 and 2050, with investment in new resources (nuclear, VRE, coal, NG combined cycle gas turbine (CCGT), NG open cycle gas turbine (OCGT), battery energy storage and transmission) considered in the last three periods. The key cost assumptions of the generation and storage resources are summarized in Table \ref{table:capcost}, where we specifically account for technology investment costs specific to the Indian context. For example, based on data from IEA \cite{IEAWEO2020}, we derate U.S. centric capital cost projections \cite{ATB2020} of wind and solar by approximately 70\% and 50\% to account for historical differences in capital cost between U.S. and India. We characterize 2020 power generation capacity as well as their operational flexibility based on the documentation of the Regional Energy Deployment Model-India \cite{NREL2020} as well as prior studies \cite{Rudnick2019} (see supplementary Table \ref{table:reg_thermal} and \ref{table:ntl_thermal}). For simplicity, we do not distinguish between the operational characteristics (e.g. minimum stable power and heat rate) of supercritical and subcritical coal power plant resources within a zone, but distinguish between heat rates of existing thermal power plants across zones (see Supplementary Table \ref{table:reg_thermal}). Major system assumptions including fuel costs and value of lost load are denoted in Supplementary Table \ref{table:sys}. Existing hydro power plants are classified as either reservoirs plants that can flexibly adjust their output vs. run-of-river resource that do not have any flexibility in their output and hence are treated as must-run resources (see Supplementary Table \ref{table:hydro}). The hourly inflows, reservoir capacity for hydro power generation are derived from a prior study \cite{Rudnick2019}. The model incorporates both lifetime-based and economic-based retirement of generation and storage resources. For existing resources, particularly coal and NG, we estimate cumulative lifetime based retirements by 2030, 2040 and 2050 by zone that represent a minimum amount of capacity to be retired by those time frames, based on data from \cite{NREL2020} (see Supplementary Table \ref{table:reg_thermal}). Because of the assumption of perfect foresight of future technology cost, VRE resource availability, demand, and policy, the model strategically may choose to retire more than the prescribed minimum capacity if it can lead to reduction in the total system cost over the modeling horizon.

\subsection*{Renewable resource supply curves}\label{method:resource}
Similar to other power system planning studies \cite{Brown2020, NREL2020, HE201674}, GenX uses supply curves to model the investment in VRE resources that account for variation in the VRE resource in terms of resource quality, interconnection cost and total deployable capacity within each zone. This supply curve is developed starting from the spatially-resolved (10 km$^2$) wind and solar resource data for 2014, available from the Renewable Energy Potential Model (reV) \cite{osti_1563140} using a sequence of steps, described in detail in Supplementary Fig. \ref{fig:flowchart_SI}. The resulting spatially-resolved capacity potential for wind and solar PV are illustrated in Supplementary Fig. \ref{fig:19_SI}, where each site is associated with a unique interconnection cost and hourly capacity factor (CF) profile. To simplify the representation of resource variation in GenX, we aggregate PV and wind resource in each zone into 3 bins that are generated by clustering sites (using k-means) based on their levelized cost of energy (LCOE). Here, the LCOE for each site is computed using site-specific CF and interconnection costs as well as capital costs and Fixed O\&M costs. The resulting parameter inputs for each resource bin per zone in the GenX model are summarized in Table \ref{table:vre} and include : a) hourly CF, computed as the weighted average CF for sites within each bin, where the weights correspond to the available area for VRE deployment associated with that site, b) total available capacity per bin, computed based on $32 MW/km^2$ and $4 MW/km^2$ for spatial density of solar and wind resources respectively \cite{NREL2020} and c) weighted average annualized interconnection cost associated with each bin.

In addition, we also impose installation limits for total wind and solar PV capacity deployed per investment period to account for potential constraints owing to supply-chain and labor resource limits. These constraints are derived based on fitting a Gompertz function growth curve to trends in VRE capacity deployments seen in the Chinese context, with further explanation provided in the Supplementary Information (see Supplementary Table \ref{table:gompertz}). Table \ref{table:caplimit} highlights the imposed decadal VRE installation limits. Supplementary Fig. \ref{fig:10_SI} demonstrates the impact of alternative assumptions about the installation limits (0.5X of the reference case and no installation limits) on generation capacity. Fig. \ref{fig:10_SI} highlights that installation limits primarily impact VRE deployment, mainly wind, in 2030 and 2040 but are less impactful in 2050 in the reference case, where value decline in VRE generation is major driver for capacity installation decisions.

\subsection*{System cost of electricity generation expansion calculation}\label{method:cost}
System average cost of electricity (SCOE), defined in Supplementary Eqn. \ref{eqn:10} is often used to to quantify the cost impacts of various technology and policy drivers. In the context of multi-period investment planning model, we define SCOE for each modeled period as the ratio of total annual system cost for the year divided by the total demand served in that modeled period. Total annual system cost includes operating cost, both fixed and variable, and annualized investment cost for the period. The latter includes investment cost of: a) resources deployed in the current period and b) resource invested in prior periods that have not yet reached their modeled lifetime and hence are accruing fixed costs related to their investment. We do not include any investment costs associated with existing generation or transmission assets as of 2020, but consider fixed operating costs for existing generation. Operational costs are calculated for the model period only based on scaling up the hourly operation costs for the modeled 20 representative weeks using their hourly weights. Since total system cost does not include unpaid investments costs of existing generation and transmission, the SCOE estimated in this study are not reflective of electricity prices for a given case but are indicative of the cost of generation expansion and are thus used for comparison of case results. Complete SCOE formulas are provided in the Supporting Information (SI).

\section*{Results}
\subsection*{Reference case}
We first evaluate model outcomes for a reference case that provides an internally consistent point of comparison to explore the impact of alternative technological and policy assumptions. For this study, the reference case is defined based on (see Table \ref{table:baseline}): a) electricity demand projections using baseline AC efficiency and evening EV charging assumptions \cite{Barbar2021}, b) technology capital cost following the "reference" trajectory, adapted from \cite{ATB2020}, in (see Table \ref{table:capcost}), c) decadal VRE installation limits derived from VRE installation trends in the Chinese context (see Methods and Supplementary Table \ref{table:caplimit}) d) NG prices held constant at \$11/MMBtu throughout the model horizon and e) no carbon policy. 

\begin{figure}[!ht]
\centering
 \includegraphics[width=15cm,height=15cm, keepaspectratio]{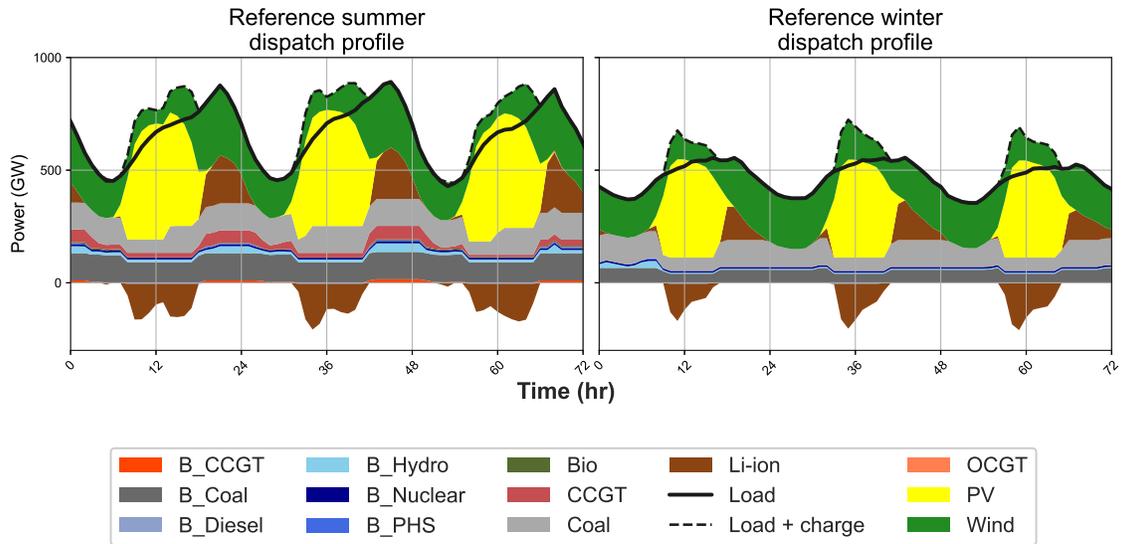}
  \caption{Hourly generation dispatch and load profile for three days during summer (left) and winter (right) periods for 2050. Model outcomes based on reference case as defined in Table \ref{table:baseline}. Storage charging is shown in the "Load + charge" curve as well as by the negative generation for storage. Technology names and their respective abbreviations in Supplementary Table \ref{table:abr}}
   \label{fig:profile}
\end{figure}

In the reference case, we find 362 GW of VRE capacity by 2030, corresponding to an annual average installation rate that is 3.7 times the average capacity additions in 2010-2019 (see Supplementary Fig. \ref{fig:10_SI}). Due to disparities in VRE resource quality and land availability, VRE generation capacity is predominantly deployed in the Southern and Western regions (see Supplementary Fig. \ref{fig:17_SI}). This is accompanied by deployment of 57 GW of new coal capacity by 2030, that along with available thermal and hydro resources, is operated flexibly to integrate the installed VRE generation with <5\% VRE curtailment (see Supplementary Fig. \ref{fig:10_SI}), as illustrated in Fig. \ref{fig:profile} and Supplementary Fig. \ref{fig:13_4_SI}. New coal capacity is predominantly installed in later periods and is concentrated in the Northern region, which has the second highest electricity demand in 2020 (427 TWh) and a 5.7\% projected growth rate (see Supplementary Fig. \ref{fig:17_SI}). Consequently, by 2050, the Northern region holds 46\% of the national coal capacity in the reference case. The role for new thermal generation, mostly coal, only becomes important as demand increases further by mid-century and VRE growth plateaus due to its decline in value with increasing penetration \cite{ATB2020}. In generation terms, this means that India could see VRE contributing over 59-66\% (see Supplementary Fig. \ref{fig:10_SI}) of annual generation in 2050, depending on the annual VRE installation rate, which we assume is limited in the reference case (see Supplementary Table \ref{table:caplimit}). Under the reference case, annual CO$_2$ emissions decline by 20\% from 2020 to 2040, but then rebounding by 48\% from the 2020 level as demand increases further by 2050. 

\begin{figure}[!ht]
\centering
 \includegraphics[width=15cm,height=15cm, keepaspectratio]{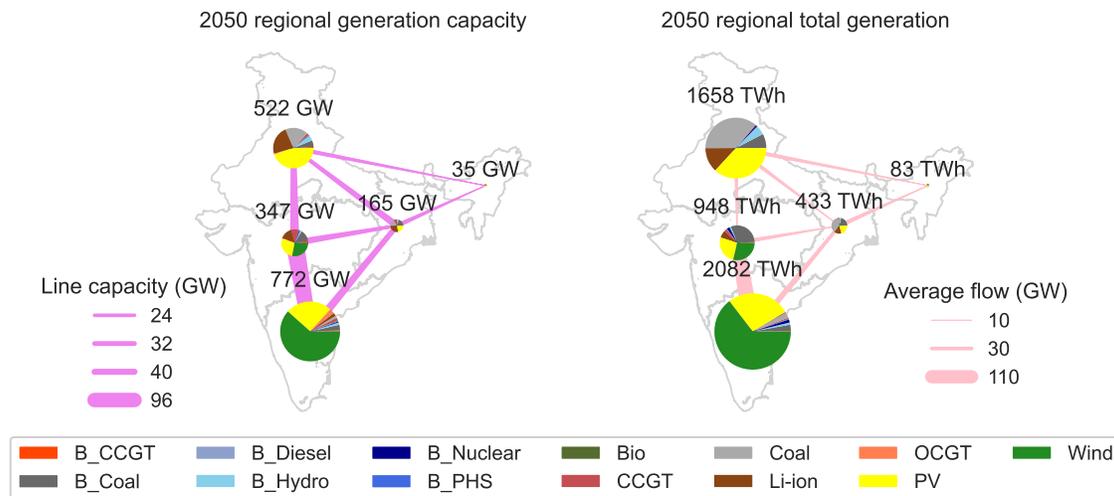}
  \caption{Regional distribution in generation and storage power capacity as well utilization trends for 2050. Regional transmission transfer capacity and its average utilization is shown.}
   \label{fig:map}
\end{figure}

Li-ion battery storage is not found to be cost-competitive until 2040. Deployment is mainly in regions with high solar PV penetration as seen in Fig. \ref{fig:map} (e.g., North), due to an increasingly stronger "duck" curve\cite{DUCK} resulting from rising solar output and rising evening demand (see Supplementary Fig. \ref{fig:13_4_SI}). Battery storage is dispatched to meet evening peak demand (see Fig. \ref{fig:profile}), with an average storage duration (i.e. total installed energy capacity divided by the total installed power per modeling period) less than 4 and a half hours in 2050. The abundant VRE resources in the Western and Southern regions and high demand in the Northern region also leads to transmission expansion in the South-West-North corridor of 77 GW by 2050, which corresponds to a 65\% increase in the transfer capacity relative to 2020 as illustrated in Fig. \ref{fig:map}. 

\subsection*{Scenario analysis spanning supply, demand drivers and policy}
We evaluated several alternative scenarios to systematically quantify the impact of various factors on electricity system evolution in the Indian context. These include: 1) high AC efficiency and low implied space cooling demand (see Methods), 2) alternative (morning, day time) EV charging schedules, 3) low capital cost for grid-scale Li-ion energy storage (see Table \ref{table:capcost}), 4) deployment of DLS storage 4) low NG prices (\$8/MMBtu), and 5) a moderate CO$_2$ policy starting at \$20/tonne and increasing to approach \$50/tonne by 2050 (see Supplementary Table \ref{table:co2}). Table \ref{table:baseline} summarizes the scenario names (columns) and their definition along various dimensions (rows). 

%In addition to demand from ACs and EV charging, for which alternative assumptions are described earlier (see Methods), grid-scale battery storage capital costs are an important parameter impacting system outcomes. While the reference case assumes mid cost projections available in the U.S. context \cite{ATB2020}, the low cost storage capital cost scenario follows the low-cost projections from the same reference. The low-cost storage scenario may be more plausible outcome for India, if storage follows trends similar to VRE in terms of cost differences between U.S. and India. Storage can not only be useful at the transmission system but also useful to mitigate peak demand in distribution systems and defer costly network upgrades, which is a particularly acute problem for loaded urban distribution systems in megacities in fast-growing EMDE countries such as India. The DLS storage sensitivity accounts for the aggregated impact of distribution level storage deployment in mega cities in the Indian context \cite{BarbarRO2021}. While gas price are relatively high in India, we look a scenario where gas price regresses to \$8/MMBtu which is the global average over the past decade \cite{stlouisfed}. Finally, even though India does not currently have a carbon policy, we investigate a CO$_2$ policy in the form of a carbon price of 20\$/tonne in 2030, 33\$/tonne in 2040 and 53\$/tonne in 2050.

\begin{table}[!ht]
\centering
\caption{Scenario definition (see Table \ref{table:capcost} for detailed costs)}
\scalebox{0.65}{
\begin{tabularx}{26cm}{l|CC|CCC|CC|C|CC|CC}
\hline
\textbf{Parameter} & \multicolumn{2}{c|}{\textbf{AC}} & \multicolumn{3}{c|}{\textbf{EV}} & \multicolumn{2}{c|}{\textbf{Storage}} & \textbf{DLS} & \multicolumn{2}{c|}{\textbf{Gas price}} & \multicolumn{2}{c}{\textbf{CO$_2$ Policy}} \\
\hline
Scenario & Baseline & High efficiency & Evening & Morning & Day & Reference & Low-cost &  & \$11/MMBtu  & \$8/MMBtu  & None &  2030:\$20/t 2050:\$52/t, \\
\hline
\hline
Reference & X &  & X &  &  & X &  &  &  &  & X &  \\
\hline
High AC efficiency &  & X & X &  &  &  &  &  &  &  & X &  \\
\hline
Low-cost & X &  & X &  &  &  & X &  &  &  & X &  \\
\hline
Low gas price & X &  & X &  &  & X &  &  &  & X & X &  \\
\hline
Morning charge & X &  &  & X &  & X &  &  & X &  & X &  \\
\hline
Day charge & X &  &  &  & X & X &  &  & X &  & X &  \\
\hline
Carbon price & X &  & X &  &  & X &  &  & X &  &  & X \\
\hline
DLS & X &  & X &  &  & X &  & X & X &  & X &  \\
\hline
High AC efficiency & \multicolumn{1}{c}{\multirow{2}{*}{}} & \multicolumn{1}{c|}{\multirow{2}{*}{X}} & \multicolumn{1}{c}{\multirow{2}{*}{X}} & \multicolumn{1}{c}{\multirow{2}{*}{}} & \multicolumn{1}{c|}{\multirow{2}{*}{}} & \multicolumn{1}{c}{\multirow{2}{*}{}} & \multicolumn{1}{c|}{\multirow{2}{*}{X}} & \multicolumn{1}{c|}{\multirow{2}{*}{}} & \multicolumn{1}{c}{\multirow{2}{*}{X}} & \multicolumn{1}{c|}{\multirow{2}{*}{}} & \multicolumn{1}{c}{\multirow{2}{*}{X}} & \multicolumn{1}{c}{\multirow{2}{*}{}} \\
-low cost & \multicolumn{1}{c}{} & \multicolumn{1}{c|}{} & \multicolumn{1}{c}{} & \multicolumn{1}{c}{} & \multicolumn{1}{c|}{} & \multicolumn{1}{c}{} & \multicolumn{1}{c|}{} & \multicolumn{1}{c|}{} & \multicolumn{1}{c}{} & \multicolumn{1}{c|}{} & \multicolumn{1}{c}{} & \multicolumn{1}{c}{} \\
\hline
High AC efficiency & \multicolumn{1}{c}{\multirow{3}{*}{}} & \multicolumn{1}{c|}{\multirow{3}{*}{X}} & \multicolumn{1}{c}{\multirow{3}{*}{X}} & \multicolumn{1}{c}{\multirow{3}{*}{}} & \multicolumn{1}{c|}{\multirow{3}{*}{}} & \multicolumn{1}{c}{\multirow{3}{*}{}} & \multicolumn{1}{c|}{\multirow{3}{*}{X}} & \multicolumn{1}{c|}{\multirow{3}{*}{}} & \multicolumn{1}{c}{\multirow{3}{*}{X}} & \multicolumn{1}{c|}{\multirow{3}{*}{}} & \multicolumn{1}{c}{\multirow{3}{*}{}} & \multicolumn{1}{c}{\multirow{3}{*}{X}} \\
-low cost & \multicolumn{1}{c}{} & \multicolumn{1}{c|}{} & \multicolumn{1}{c}{} & \multicolumn{1}{c}{} & \multicolumn{1}{c|}{} & \multicolumn{1}{c}{} & \multicolumn{1}{c|}{} & \multicolumn{1}{c|}{} & \multicolumn{1}{c}{} & \multicolumn{1}{c|}{} & \multicolumn{1}{c}{} & \multicolumn{1}{c}{} \\
-carbon price & \multicolumn{1}{c}{} & \multicolumn{1}{c|}{} & \multicolumn{1}{c}{} & \multicolumn{1}{c}{} & \multicolumn{1}{c|}{} & \multicolumn{1}{c}{} & \multicolumn{1}{c|}{} & \multicolumn{1}{c|}{} & \multicolumn{1}{c}{} & \multicolumn{1}{c|}{} & \multicolumn{1}{c}{} & \multicolumn{1}{c}{} \\
\hline
DLS and low-cost & X &  & X &  &  &  & X & X & X &  & X & \\
\hline
\end{tabularx}}
\label{table:baseline}
\end{table}

% Figure 2 - % Geographical trends
% Map shown capacity and generation trends in 2030 as in MP deck

% Takeaways the north has more demand so it will have more coal since transmission from south to north cannot be justified with building more renewables due to land constraint and demand growth - compare demand growth per region (chart annual demand by zone by time frame). Storage is complimentary to solar. check renewables capacity limitations per region. storage capacity to power - duration of storage on a regional basis. transmission capacity south to west expansion to evacuate wind to the north.

\subsection*{Impact of key supply and demand-side drivers}
\begin{figure}[!ht]
\centering
 \includegraphics[width=15cm,height=15cm, keepaspectratio]{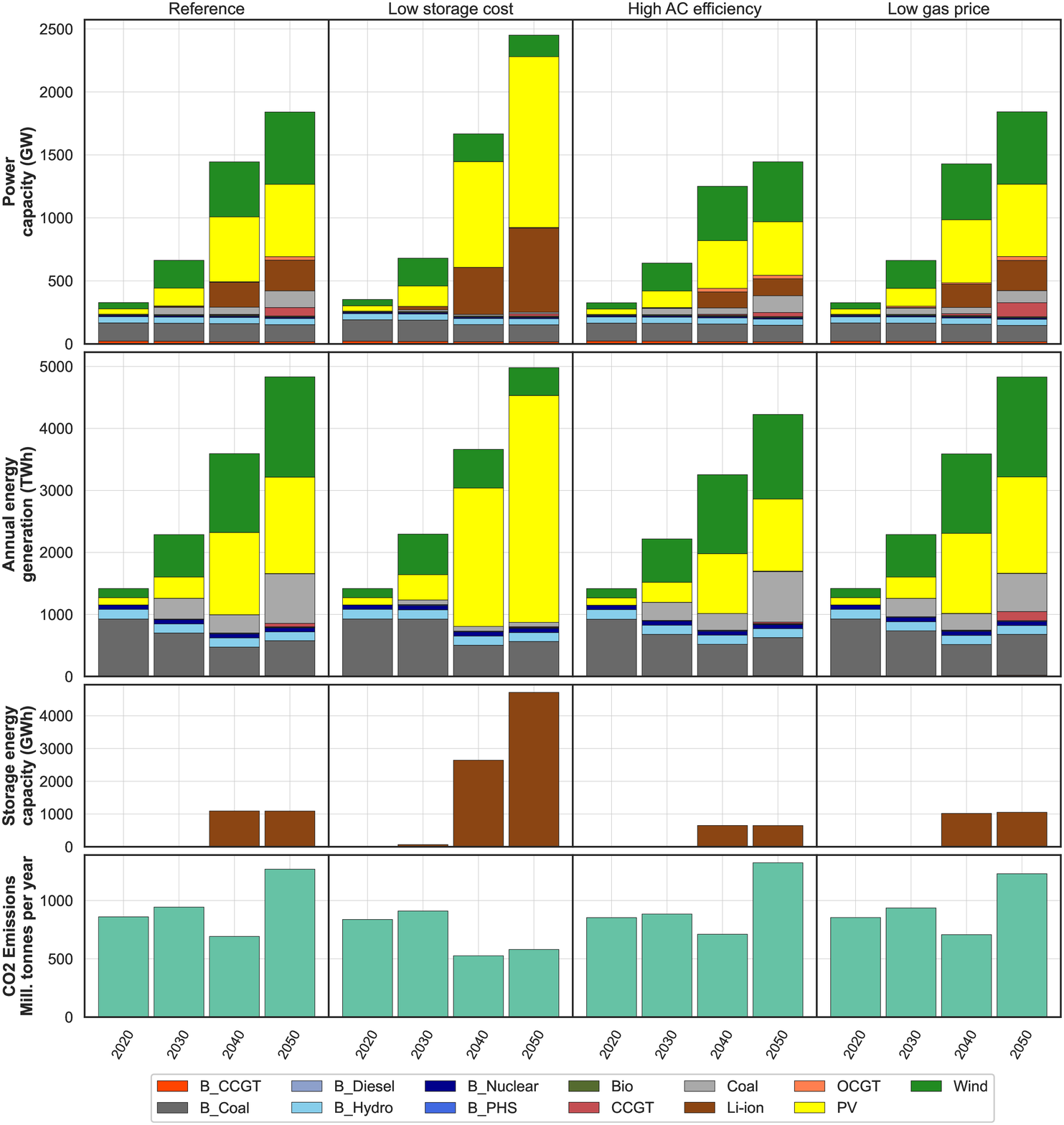}
  \caption{Installed capacity (1$^{st}$ row), annual energy generation (2$^{nd}$ row), storage energy capacity (3$^{rd}$ row) and annual CO$_2$ emissions (4$^th$ row) for reference case (1$^{st}$ column), as well as cases with alternative assumptions for battery storage capital cost (2$^{nd}$ column), high AC efficiency (3$^{rd}$ column) and gas prices (4$^{th}$ column). Detailed assumptions for each case are provided in Table \ref{table:demand} and Table \ref{table:capcost}}
   \label{fig:reference}
\end{figure}

Fig. \ref{fig:reference} and Supplementary Fig. \ref{fig:ev} highlight the impact of four main technological parameters spanning demand and supply, on the least-cost evolution of the Indian grid. Specifically, we evaluate model outcomes based on alternative assumptions for each parameter and compare them to the reference case. 
% Figure 3 - Impact of AC efficiency:
%figure showing 2020-2050 power capacity (row 1), generation (row 2), emissions (row 3) for 
%% - baseline scenario (column 1)
%% - low cost storage (column 2)
%% - high AC efficiency (column 3)
%%  - low cost storage + high AC (column 4) 
%%%  Text - discuss storage energy capacity, transmission expansion trends and put them in the SI

%\subsubsection*{Storage cost and NG prices}
% storage takeaways: storage drops emissions by a factor of two in low cost scenario - more hopeful realization, impact of storage deploys more renewables
% storage duration notable change across scenarios account for discharge efficiency in the text - if there is anything notable.
While the reference case assumes mid cost projections for the U.S. context \cite{ATB2020}, the low-cost storage scenario follows the low-cost projections from the same reference. The low-cost storage scenario may be more plausible outcome for India, if storage follows trends similar to VRE in terms of cost differences between U.S. and India. In the low battery storage cost case\cite{ATB2020}, storage power and energy capacity increases by 424 GW (174\% increase) and 3,625 GWh (332\% increase), which enables 30\% more VRE generation in 2050 compared to the reference case. This results in 54\% lower annual CO$_2$ emissions compared to the reference case in 2050, the highest reduction among the considered parameter sensitivities (3\% reduction for low gas price case and 4\% increase for the high AC efficiency case in Fig. \ref{fig:reference}). This is largely due to the increased competitiveness of VRE, especially solar, which reduces new coal installations by 91\% compared to the reference case by 2050. Low-cost storage reinforces the deployment of VRE, and leads to VRE supplying 65\% of annual generation in 2050, accompanied by transmission level storage of average duration under 7 hours. This is achieved with system average cost of electricity in 2040 and 2050 being 22\% and 39\% lower than the reference case (see Fig. \ref{fig:cost}) and 92\% less transmission expansion capacity by 2050 compared to the reference case (see Supplementary Fig. \ref{fig:9_SI}). We note that among the supply and demand drivers considered, storage cost are the important factor affecting annual CO$_2$ emissions in 2050. Moreover, low-cost storage eliminates the need to build new coal capacity particularly in 2040 and 2050 where we note mass deployment of storage as an enabling additional VRE generation.

Under the reference case, AC demand growth is projected to contribute 43\% to peak demand in 2050, thereby creating the need for peaking generation capacity. Combined cycle gas turbines (CCGT) and open cycle gas turbines (OCGT) are best fit for peaking generation due to their greater operational flexibility compared to coal power plants and lower capital costs (see Table \ref{table:reg_thermal} and Table \ref{table:ntl_thermal}). Moreover, given the relatively high cost of NG fuel vs. coal (see Supplementary Table \ref{table:sys}), NG generation capacity is deployed but utilized sparingly, with annual capacity utilization for CCGT and OCGT plants at 5\% and 3\%, respectively, in 2040 under the reference case (see Supplementary Fig. \ref{fig:11_SI}). Because of this, the deployment of new NG generation capacity is closely tied to AC demand growth, with the high AC efficiency scenario virtually eliminating the need for new NG capacity (see Fig.\ref{fig:reference}, column 3 vs. 4) in 2050. At the same time, low NG prices improve the economic viability of NG generation, leading to higher CCGT capacity deployment and utilization vs. the reference case (see Supplementary Fig. \ref{fig:11_SI}) in 2050 that also reduces new coal capacity deployment by 28\%. Low NG prices erode coal generation without significantly changing VRE deployment, owing to the operational flexibility of gas generation that, along with storage, complements integration of VRE generation at similar level of curtailment (7\% in 2040). This explains why the low NG price case has 3\% lower annual CO$_2$ emissions in 2050 compared to the reference case. It is important to note that India is both gas resource and infrastructure constrained, and the modeled low NG price case reflects the global average liquefied natural gas price over the past decade \cite{stlouisfed}. However, it  does not consider the cost of the required NG transmission infrastructure (pipelines, compressor stations), nor the implied costs that may obtain from siting constraints.

% low gas price: mechanism to get to the same impact as high ac efficiency versus efficient ac emissions achievement is the same but through gas allowing renewables to be more deployable LNG is likely to grow due to peakiness if gas prices go down should they get gas from somewhere. low capacity factor.

We use a bottom-up demand forecasting model \cite{Barbar2021} to evaluate electricity demand under a high AC efficiency scenario in which India's average SEER rating trails the efficient global weighted average (8.5) by 15\% as opposed to being 36\% behind in the baseline global weighted average SEER (6.2) in 2050 \cite{IEAcooling18}. Fig. \ref{fig:reference} highlights the supply-side impacts of the modeled AC demand, where we account for regional disparities in AC adoption, which depend on climate and population size, giving rise to significant variation in estimated electricity demand for space cooling \cite{Barbar2021}. The high AC efficiency case in Fig. \ref{fig:reference} shows a 22\% and 13\% decrease in installed capacity and generation, respectively. AC demand accounts for over 40\% of peak demand in summer 2050, occurring during evening hours (8 PM to 12 AM), under the reference case. However, in the high AC efficiency case, AC demand accounts for less than 20\% in the high AC efficiency case in summer 2050 (Table \ref{table:demand}). In addition to reducing capacity and generation requirements, the reduction in AC demand also results in a flatter demand profile that has two further supply-side impacts: a) it reduces the value for peaking generation provided by NG power plants and battery storage and b) reduces the value of solar generation in serving demand in the day time as well as indirectly meeting evening peak demand via battery storage (see Supplementary Fig. \ref{fig:13_5_SI}). The impact of high AC efficiency on CO$_2$ emissions is most notable in 2030 when storage is not yet cost competitive. High AC efficiency is responsible for 7\% annual CO$_2$ emissions reduction in 2030 when compared to the reference case and slightly lower system average cost of electricity (Fig. \ref{fig:cost}). Absent a carbon emissions constraint or low-cost storage, a flatter demand profile, however, leaves more room for baseload generation provided by existing coal and further investment in new coal generation in 2040 and 2050. Supplementary Fig. \ref{fig:13_4_SI} compares the grid dispatch over a summer week for the reference case and high AC efficiency case, which suggests that peak demand is a key driver of value for solar and by association storage in the system. In a regime when peak demand is reduced compared to the reference case and with the existing coal assets that India currently holds in its generation portfolio, the role of storage is not as significant and therefore less solar is built to charge storage. When less solar is built out, the optimization model pivots to more coal generation that is not affected by seasonality. This explain why high AC efficiency does not improve grid CO$_2$ emissions intensity of the subsequent modeled periods (see Supplementary Fig. \ref{fig:9_SI}). We further note that high AC efficiency on its own does not contribute to cumulative (i.e. summed over all model periods) emissions reduction. Cumulative emissions for the reference case is 3,766 million tonnes while the high AC efficiency case results in 3,772 million tonnes of CO$_2$ emissions.  Therefore, while implementing demand-side efficiencies has clear positive outcomes, both in terms of cost and CO$_2$ emissions, in the short run (2030), a sustainable and continuous supply-side effort needs to complement it in the long-term to ensure emissions mitigation and cost-effectiveness. 

Another possible demand driver is EV charging, which unlike AC use, also has the potential to offer flexibility to the system \cite{NRELEFS2018}. The reference case assumes EV charging predominantly takes place during evening hours (7 PM to 12 AM), which is generally consistent with residential EV charging schemes \cite{berkeley}. If instead, EV charging is predominantly shifted to morning hours (i.e., 5 AM to 10 AM), reflective of mixed charging infrastructure deployment, it reduces the contribution of EV charging to peak demand to 6\% vs. 10\% in the reference case \cite{Barbar2021}. Supplementary Fig. \ref{fig:ev} (top) quantifies the grid impacts of morning relative to evening EV charging (reference case). As EV demand grows over time, charging demand during the evening hours can be met via short-duration battery storage that is charged during day time hours when solar generation is prevalent. However, if EV demand were to occur in the early morning hours, Supplementary Fig. \ref{fig:ev} (bottom) highlights that deployment of overnight energy storage to discharge in the morning is not cost-effective and instead coal deployment is favored. The net impact is that morning EV charging schemes favor coal deployment over VRE and storage and result in a 2\% higher system average cost of electricity in the three investment periods as well as 3\% higher annual CO$_2$ emissions in 2050. As one might expect, aligning EV charging with periods of high solar irradiation gives rise to more solar with little or no storage, and no new coal generation. In this case, installed capacity of coal is reduced by 2\% while solar and wind capacity is increased by 2\%, and resulting in a less than 2\% reduction in annual CO$_2$ emissions.

% Fig 6 EV charging scheme comparison
%  If I read this correctly, workplace charging favors solar over wind, that also mean increase need for battery storage to manage even peaks and by 2050 could marginally impact emissiosn.
% Sentence for conclusion - Thus, the choice of an EV charging scheme can have a small but meaningful impact on peak generation and emissions. 
% versus AC efficiency a more realizable scenario. but you are flattening the load curve so less need for peaking plants and more baseload plants, coal more utilized Emissions intensity increases (SI emissions intensity plot - capacity factor plot by generation (group generations types from various scenario in one panel)

%Low energy storage combined with low gas prices costs could significantly reduce new coal capacity additions even without a carbon price, leading to lower emissions in 2050 than in the reference scenario.

%Storage deployment is closely tied with cooling demand growth. Improved air conditioning efficiency directly impacts midnight peak load and reduces need for energy storage.

\subsection*{Impact of distribution-level storage (DLS)}
The projected growth in peak demand can not only drive investments in centralized generation and transmission capacity, but also at the distribution network level. For the latter, battery storage is increasingly viewed as viable non-wire alternative (NWA) network relief mechanism that can allow for deferring network upgrades with large financial impacts due to high cost of capital \cite{6450147, KIM2017918, BarbarRO2021}. Deployment of battery storage to partially offset peak demand within the distribution system modifies the demand profile seen by the transmission system owing to timing and duration of battery charging and discharging.  We compute such a "transmission level" demand and the accompanying DLS deployment based on outcomes from modeling the operation and sizing of storage in urban distribution feeders in the Indian context using real options and linear optimization framework, described elsewhere \cite{BarbarRO2021} (See Supplementary Method on DLS analysis). Because DLS is only deployed when network deferrals are economic \cite{Evans2020}, i.e. the present value of investments in battery storage is less than that of investments in network upgrades, their impact on the transmission system can be captured via the modified transmission level demand (demand + DLS charging - DLS discharging, see Supplementary Fig. \ref{fig:14_SI}) without representing DLS's capital or operating cost. Across the regions, cost-optimal DLS sizing points to an average storage duration (ratio of energy capacity to power capacity) of 2 to 6 hours that is consistent with the duration of the overloading peak demand as well as available off-peak charging hours without violating distribution network capacity constraints. In 2030, a total of 93 peak hours were shaved with deployment of DLS storage of 29 GWh nationally for the reference case demand scenario.
%We compute such a "transmission level" regional demand for two cases with DLS deployment: 1) the reference case and 2) the low storage cost case.
\begin{figure}[!ht]
\centering
 \includegraphics[width=12cm,height=12cm, keepaspectratio]{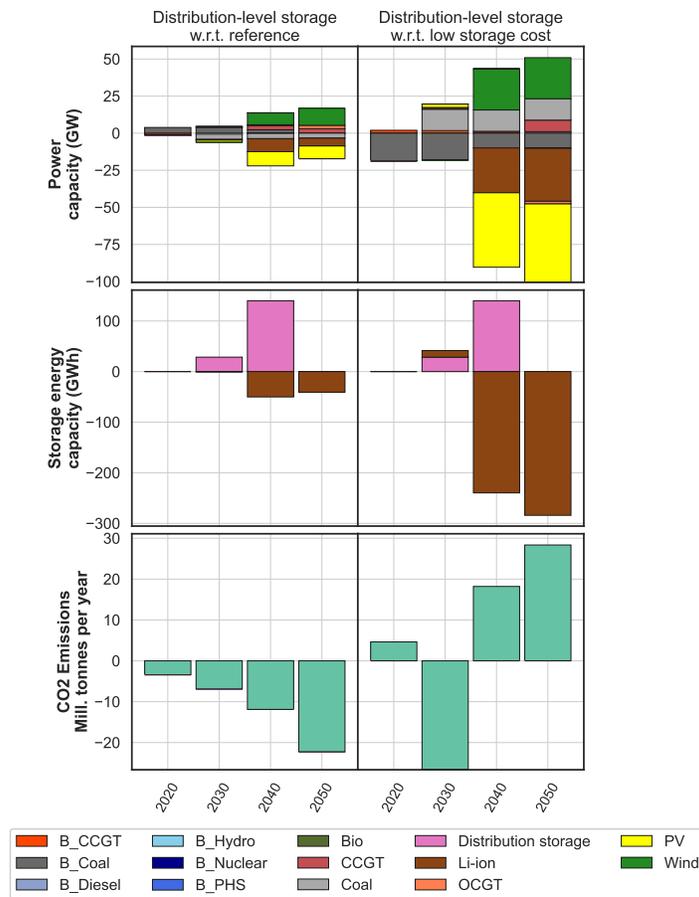}
  \caption{Impact of distribution-level storage deployment on dispatched generation (1$^{st}$ row), installed storage energy capacity (2$^{nd}$ row) and annual CO$_2$ emissions (3$^{rd}$ row) under the reference case (1$^{st}$ column) and the low-cost storage (2$^{nd}$ column) case.}
   \label{fig:dls}
\end{figure}

From the transmission system perspective, DLS, when deployed, reduces peak demand that occurs during evening hours, while increasing demand by charging during off-peak hours. Supplementary Fig. \ref{fig:dls-impact} highlights the temporal changes in transmission level demand from DLS discharging during evening hours and charging during earlier hours in the day (7 AM to 11 AM) when solar availability is not maximized. DLS operation aims to minimize peak demand and network upgrades and thus spreads out the charging over several hours rather than maximize charging during periods of abundant low-marginal cost supply from resources like solar. Consequently, Figure \ref{fig:dls} shows that DLS deployment tends to shift the installed capacity mix to favor wind that has high capacity factors at night and early morning hours as well as coal resources over solar PV and battery storage. By 2050, demand growth has sufficiently materialized and DLS is no longer cost-effective as an alternative to network upgrades, and is consequently retired. The correlation between storage and peak demand is most pronounced under the low-cost storage case with DLS deployment (Fig. \ref{fig:dls} 2$^{nd}$ column), where the DLS enabled peak shifting has a trickle down effect on the generation design, since less storage is needed for peak hours discharging and therefore less solar capacity is installed to charge up the storage. Since demand is being met by either alternative VRE with less intra-day variability (i.g. wind) or coal, the role for solar + storage at scale is thus reduced with the presence of DLS, even under the assumption of low-cost storage. Over 90 TWh of additional wind capacity is installed with respect to the low-cost storage case (see Fig. \ref{fig:reference} 2$^{nd}$ column) but also more coal since DLS results in flattening the demand profile. Overall, factors like DLS and AC efficiency improvements are demand-side peak shifting or reducing mechanisms that, depending on the cost of storage, can indirectly lead to increase (reference case) or decrease (low-cost storage) coal generation. At the distribution level, DLS or high AC efficiency are clearly a cost saving mechanism that helps distribution companies minimize capital investment\cite{1717560, 6450147}. When aggregating DLS to transmission-level for national planning considerations, the overall system cost of electricity does not improve compared to the reference case (see Supplementary Table \ref{table:dls} and accompanying discussion) and therefore necessitates further attention to trade-offs between transmission and distribution DERs \cite{Burger2019}. It should be noted the impact of DLS deployments modeled here are relatively small when compared against the impacts of AC efficiency improvements as well as the total capacity deployed on the system in the reference case (Fig. \ref{fig:reference}).

\subsection*{Technological vs. policy drivers to reduce new coal investments}
\begin{figure}[!ht]
\centering
 \includegraphics[width=15cm,height=15cm, keepaspectratio]{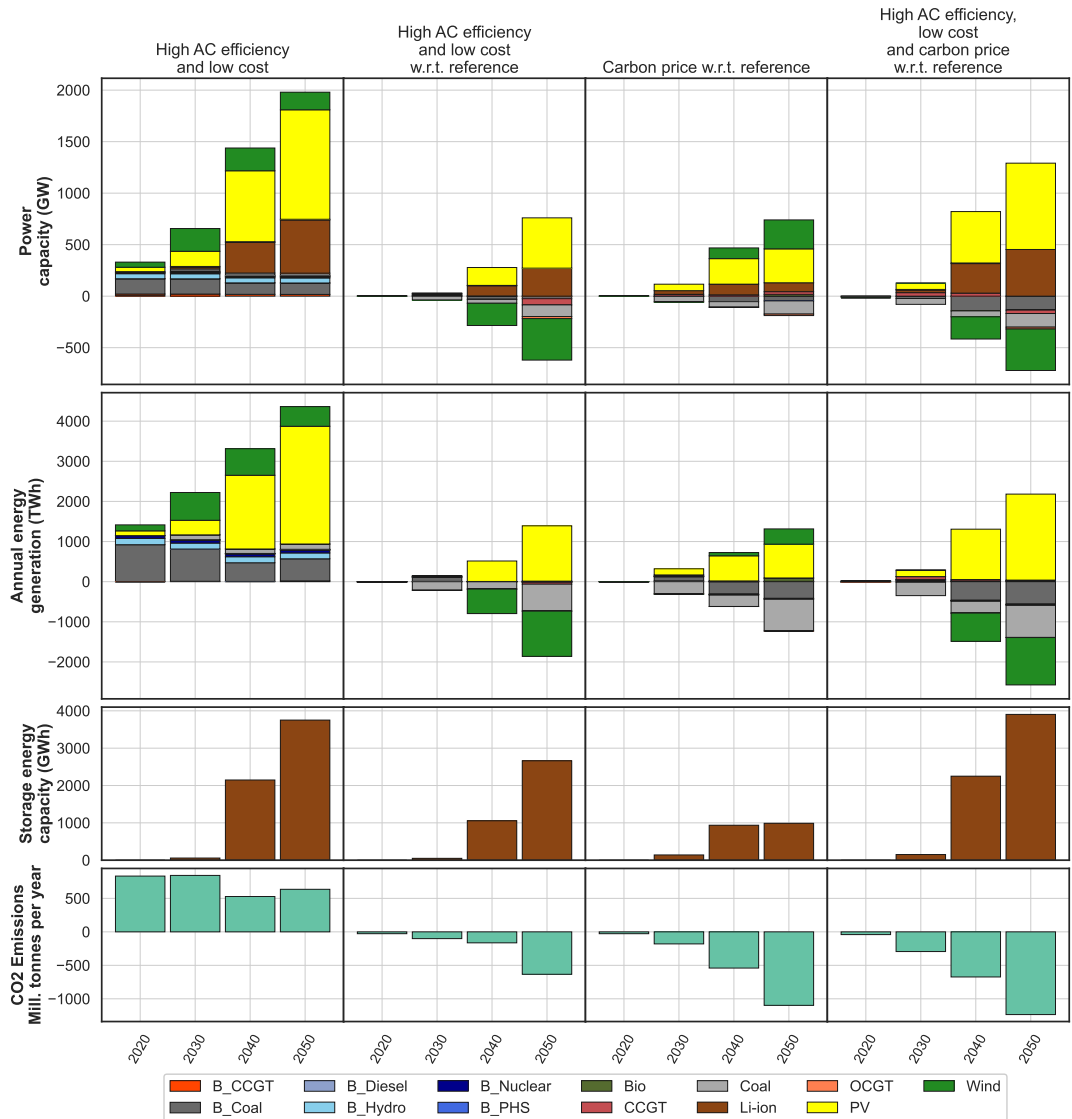}
  \caption{Model outcomes for high AC efficiency-low cost case, defined by low battery storage capital cost, high AC efficiency and low NG price (1$^{st}$ column) as well as impact of carbon price with and without scenario assumptions. Model outcomes include installed capacity (1$^{st}$ row), annual energy generation (2$^{nd}$ row), storage energy capacity (3$^{rd}$ row) and annual CO$_2$ emissions (4$^th$ row). Columns 2-4 highlight outcomes compared to the reference case for the following cases: a) high AC efficiency-low cost case (2$^{nd}$ column), b) low carbon price case, where CO$_2$ price starts at 20\$/tonne in 2030 and grows by 5\% each year (3$^{nd}$ column) and high AC efficiency-low cost + low carbon price scenario case (4$^{rd}$ column).}
   \label{fig:sensitivity}
\end{figure}

The outcomes of the individual technology cases point to possible strategies for minimizing future investment in stranded coal generation under global climate mitigation goals. This raises the question whether a combination of these approaches will be most beneficial for coal reduction. The high AC efficiency-low cost case, as defined in Fig. \ref{fig:sensitivity}, highlights the collective impact of low-cost storage, low NG prices and high AC efficiency on power system evolution, where we see the combined effects of these supply- and demand-side drivers. As discussed above, low NG prices and high AC efficiency favor fossil generation (gas and coal, respectively) over solar and battery storage to meet peak demand compared to the reference case, while low-cost storage increases deployment of solar and storage. Collectively, in the high AC efficiency-low cost case, these factors lead to a 112\% increase in need for energy storage power capacity compared to the reference case (Fig. \ref{fig:sensitivity} column 2) by 2050, while energy capacity deployment increase by 244\% in 2050 compared to the reference case (see Supplementary Fig. \ref{fig:21_SI}). The flatter demand profile (on account of high AC efficiency) and lower energy storage costs, increases duration of storage deployed as compared to the reference case as well as low storage cost case (see Supplementary Table \ref{table:bat_duration}). Overall, this case leads to a 50\% reduction in annual CO$_2$ emissions over the reference case in 2050 and the lowest cumulative emissions (i.e. sum of all modeled annual emissions) of any individual case considered here. Annual CO$_2$ emissions reductions in 2030 vs. reference case are attributed to high AC efficiency. Further down the line, emissions savings vs. the reference case are primarily attributed to low-cost storage that supports solar integration. Still, CO$_2$ emissions in 2050 are 1.3 times higher than the low-cost storage case, in part because high AC efficiency leads to a flatter load profile that reduces the value of solar relative to coal generation, all else remaining equal. Flatter demand profile also reduces the need for peaking NG generation (Fig. \ref{fig:sensitivity} column 2) compared to the reference case.

%We note a strong correlation between solar charged storage and peak AC demand: a flatter load profile will lead to more baseload generation and in the case of India this is primarily serviced by coal. The reason why we see mass deployment of storage in many of the cases is because of the projected increase in space cooling peak demand which can be supplied by short duration storage charged by cheap solar resources. Solar + storage does not prove competitive when the demand is spread out throughout the day i.g. high AC efficiency.

\begin{figure}[!ht]
\centering
 \includegraphics[width=15cm,height=15cm, keepaspectratio]{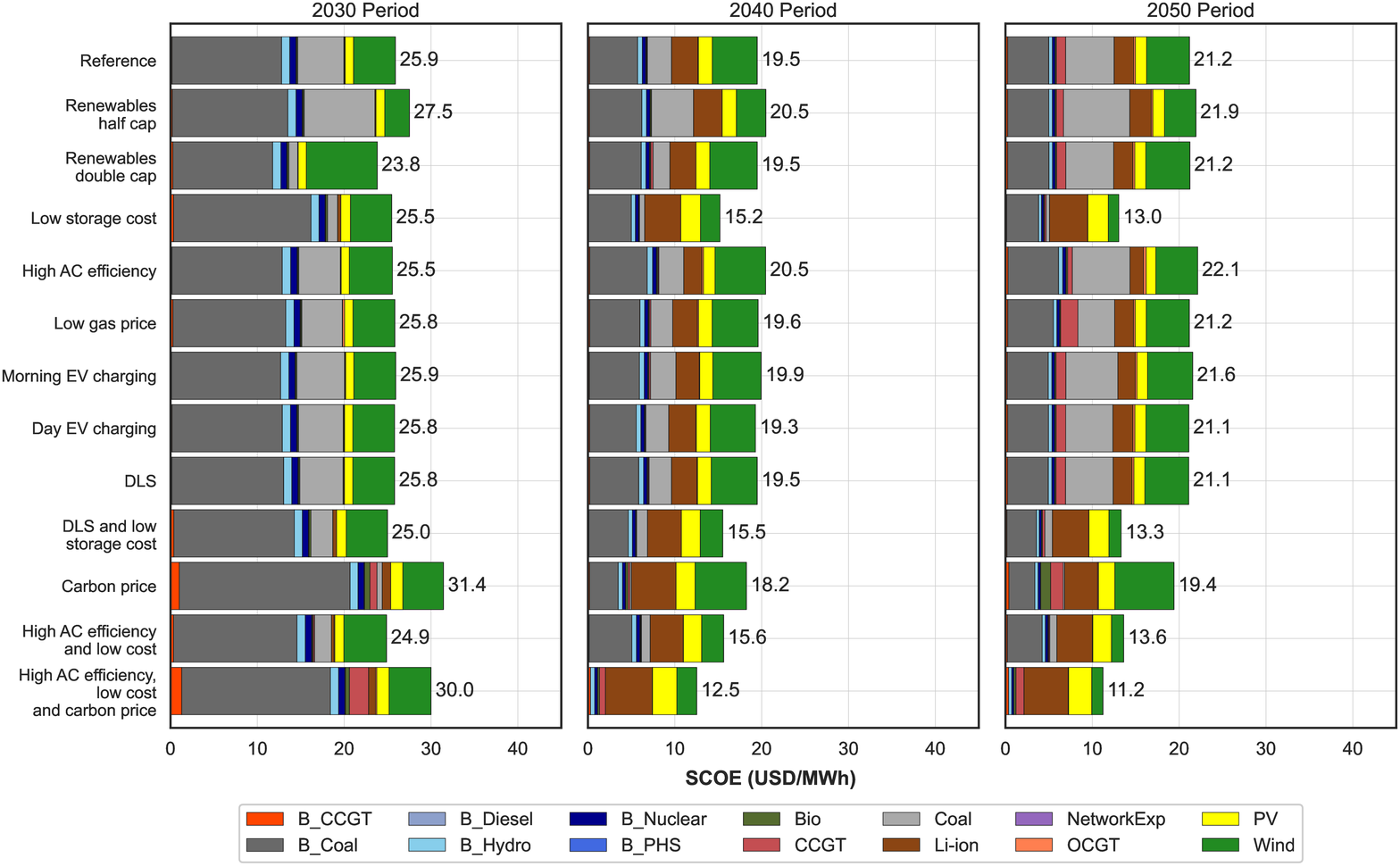}
  \caption{System average cost of electricity generation expansion (SCOE) per modeling period for a range of cases evaluated in the study. See methods and Supplementary Information for SCOE computation. Note that SCOE does not include fixed costs associated with existing generation and transmission assets in 2020.}
   \label{fig:cost}
\end{figure}

Although such a technology-focused strategy can substantially reduce new coal investment, existing coal capacity (see Supplementary Table \ref{table:reg_thermal}) remains operational up to mid-century in the high AC efficiency-low cost case and contributes 16\% of total generation even in 2050. This suggests that demand- and supply-side mechanisms to reduce emissions are insufficient for deep decarbonization of the grid, and additional policy measures may be needed. As an example policy measure, Fig. \ref{fig:sensitivity} explores the impact of a CO$_2$ price that starts at 20 \$/tonne in 2030, and increases by 5\% over time, approaching 50 \$/tonne by 2050 (see Supplementary Table \ref{table:co2}). There is precedent for carbon pricing in India. Since 2010, India has imposed a tax on coal production, which has increased steeply from INR 50 (\$0.70)/tonne coal in 2010 to INR 400 (\$5.61)/tonne coal since 2016. This practice is included in India's nationally determined contribution under the Paris agreement \cite{iisd2018}. As compared to the reference case (see Fig. \ref{fig:sensitivity}), the expectation of a rising CO$_2$ price leads to reduced utilization and early retirement of existing coal, discourages investments in new coal, and favors increased investment in low-carbon generation, mainly VRE and storage. Fig. \ref{fig:cost} highlights that the cost impacts of a carbon policy are greatest in 2030 (21\% higher system cost of electricity vs. the reference case) when existing coal supplies 30\% of annual demand. However, in 2040 and 2050, a combination of factors, including declining coal utilization, investment in VRE and storage vs. new coal reduces the system annual cost of electricity by 20\% (see Fig. \ref{fig:cost}), while reducing CO$_2$ emissions by 86\% compared to the reference case. It is evident, therefore, that the initial electricity cost increase resulting from a carbon price can be mitigated over time by combining it with technological measures of the high AC efficiency-low cost case as shown in Fig. \ref{fig:cost}, primarily attributable to the low-cost of storage. The technology + policy approach reduce costs by 36\% and 47\% compared to the reference case for 2040 and 2050 respectively, and paves the way for grid decarbonization by mid-century through reducing CO$_2$ emissions by 97\% compared to the reference case, retiring all of the existing coal in 2050 and reducing emissions intensity to 8 gCO$_2$/kWh (see Fig. \ref{fig:sensitivity} column 4).

% technology combination, much of the gain is coming from low cost storage allowing more renewables
% technology based efficiency - technology will only drive out new coal existing coal will still be economical unless a carbon price is added to retire existing coal and eliminate emissions
% policy based efficiency column 3 base + low carbon price 
% both
% potential for long duration storage with same cost assumptions for 2040 and 2050 only available there.

\section*{Discussion}
India represents one of many countries that will have to contend with electricity system CO$_2$ emissions impacts of rapidly growing electricity demand from space cooling and to a lesser extent, EV adoption, over the next 3 decades given its dependence on coal (953 GWh in 2020). Here, we present an analytical framework for evaluating the implication of such demand drivers in conjunction with other supply drivers on cost-optimal pathways to electricity system decarbonization by mid-century. We demonstrate this framework through a detailed assessment of India's electricity system, where we find, as have other recent studies \cite{Deshmukhe2008128118}, that large amounts of VRE generation, enabled by storage, are an important feature of a least-cost expansion of the electricity supply over the next 3 decades. However, growth and changing temporal patterns in electricity demand, driven by AC use, are projected to outstrip growth in cost-optimal VRE generation in our reference case and could lead to 48\% higher CO$_2$ emissions in 2050 vs. 2020 levels. NG plays a marginal role, i.e. during extreme peak demand hours when solar and storage are exhausted, in the generation portfolio of the Indian electricity system. Installation of large capacity of low utilization NG turbines will marginally reduce coal dispatch without any significant impact on emissions, as long as prevailing NG price trends continue. Moreover, relatively lower NG prices that are within expectations of long-term liquefied natural gas prices, can only partially substitute new coal capacity by NG, but cannot displace existing coal capacity.

High AC efficiency reduces total generation and storage capacity and produces short-term CO$_2$ emissions and cost reductions, but without further incentives to restrict coal generation, is not expected to contribute to long-term emissions reduction efforts. Among the technology cost scenarios we have modelled, the cost of storage has the greatest impact on long-term CO$_2$ reductions. Storage complements solar generation by time-shifting either generation (transmission level storage) or demand (DLS). We further note that even small time shifts in demand drivers can have an incremental effect at bulk power system level as seen by EV and DLS. While we show that both demand and technological measures are valuable in reducing dependence on coal and increasing VRE penetration, complementing those measures with an incremental carbon price, or an equivalent measure, is the clearest pathway for deep decarbonization of India's electricity sector. With forty Indian companies already committed to an internal carbon price \cite{cdp2017}, the policy pathway is a pragmatic and well within reach solution that could place India as a global leader in VRE and grid-scale battery storage deployment.

Although these findings are based on a study of India's electricity system, many aspects can be generalized to other parts of the world, where similar supply- and demand-side factors persist (e.g. Indonesia, Nigeria, Vietnam). As an example, India's growth due to increased cooling demand can also be seen in Sub-Saharan African countries where as in India, dependence on fossil fuels is very strong and NG availability is constrained. The analytical framework developed here can be applied to these other regions to develop holistic view for electricity system decarbonization pathways by mid-century at the global scale based on considering available technology, local resources and, practically viable policy approaches relevant to each region. 

%\subsection*{Limitations, assumptions and future work}\label{method:limit}
We note several limitations of this work. On the technology aspect, the cases evaluated do not consider deployment of certain low-carbon resources like hydro, nuclear or carbon capture and sequestration (CCS) equipped fossil-fuel power plants. Since investments in hydro and nuclear are not driven solely by economics, they are not considered in the model. Additionally, while hydro is being deployed in India recently, the expected increase in capacity is 12 GW \cite{IndiaEnergyOutlook, CEANEP2018} which is minimal compared to the projected peak demand. Similar, India's nuclear generation goals set by the Central Electricity Authority are also low, and are therefore not likely to drastically change the modeling outcomes presented here \cite{CEANEP2018}. CCS has not yet been considered in Indian national electricity plan \cite{CEANEP2018} or is considered a post 2050 technology which is out of scope of the presented results \cite{MORRIS2019170}.

We also restrict short duration battery storage technology to lithium-ion due to its rising popularity and declining costs. While lead acid batteries are presently more prevalent in India, we assume that by 2040 --- where we note mass deployment of grid scale short duration storage --- lithium-ion dominates the market. On the modeling aspect, GenX simulates grid economic dispatch which does not reflect the current structure of the electricity system dispatch in India. Additionally, we do not take into consideration administrative transmission losses, due to theft and other exogenous events, when modeling simplified regional transmission flows (see Fig. \ref{fig:map}). Finally, the resource availability maps used for VRE characterisation were processed using satellite capacity factor data \cite{michael_rossol_2021_4501717} which includes 14\% system losses, with 1.5\% corresponds to light-induced degradation \cite{pvwatts}. However, ground truth data may differ due to smog and poor air quality, particularly in case of PV. This might lower the value of PV compared to our modeling outcomes.

\bibliography{sample}

%\noindent LaTeX formats citations and references automatically using the bibliography records in your .bib file, which you can edit via the project menu. Use the cite command for an inline citation, e.g.  \cite{Hao:gidmaps:2014}.

%For data citations of datasets uploaded to e.g. \emph{figshare}, please use the \verb|howpublished| option in the bib entry to specify the platform and the link, as in the \verb|Hao:gidmaps:2014| example in the sample bibliography file.

\section*{Acknowledgements}
This research was supported by MIT Energy Initiative's Low Carbon Energy Centers and the Future of Storage Study. We thank Tata Power Delhi Distribution Limited for their cooperation and contribution to the input data, the MIT Energy Initiative Future of Storage study team members for their review of the results, Aaron Schwartz in implementing the dual dynamic programming algorithm as part of the GenX model version used in the study and, Prof. Ignacio Perez-Arriaga for his feedback on the manuscript. The authors acknowledge the MIT SuperCloud and Lincoln Laboratory Supercomputing Center for providing HPC resources that have contributed to the research results reported within this paper.

\section*{Author contributions statement}

M.B. and D.M. conceived the experiment(s), M.B. and D.M. conducted the experiment(s), M.B., D.M. and R.S. analyzed the results.  All authors reviewed the manuscript. 

\section*{Additional information}

The authors declare no competing interests.

\section*{Figures \& Tables}

\begin{table}[!ht]
\centering
\caption{Demand estimates from bottom-up forecasting model \cite{Barbar2021} that is used as inputs to the supply-side modeling. }
\begin{tabular}{ll|cccc}
\hline
\multicolumn{2}{c|}{}& \textbf{2020}& \textbf{2030}& \textbf{2040}& \textbf{2050}\\
\hline
\multicolumn{1}{c|}{Reference} & Peak demand (GW)& 197 & 347 & 626 & 901 \\
\multicolumn{1}{c|}{case} & Annual demand (TWh)& 1,421 & 2,282 & 3,523 & 4,773 \\
\hline
\multicolumn{1}{c|}{High AC} & Peak demand (GW)& 197 & 317 & 501 & 677 \\
\multicolumn{1}{c|}{efficiency case} & Annual demand (TWh)& 1,421 & 2,207 & 3,205 & 4,199 \\
\hline
\multicolumn{1}{c|}{DLS} & Peak demand (GW)& 197 & 341 & 600 & 901 \\
\multicolumn{1}{c|}{case} & Annual demand (TWh)& 1,421 & 2,282 & 3,523 & 4,773 \\
\hline
\multicolumn{1}{c|}{EV charging} & Peak demand (GW)& 197 & 345 & 624 & 897 \\
\multicolumn{1}{c|}{case} & Annual demand (TWh)& 1,421 & 2,282 & 3,523 & 4,773 \\
\hline
\hline
\multicolumn{1}{c|}{AC contribution} & Baseline efficiency & 4\%& 15\%& 32\%& 42\%\\
\cline{2-6}
 \multicolumn{1}{c|}{to peak demand} & High efficiency& 4\%& 10\%& 17\%& 19\%\\
 \hline
\multicolumn{1}{c|}{EV contribution}  & Evening charging& 1\%& 4\%& 6\%& 10\%\\
\cline{2-6}
\multicolumn{1}{c|}{to peak demand} & Morning charging& 1\%& 3\%& 5\%& 9\%\\
\hline
\end{tabular}
  \label{table:demand}
\end{table}

\begin{table}[!ht]
  \centering
  \caption{Capital cost assumptions for various resources. All costs in 2018 dollars and sourced from NREL annual technology baseline 2020 \cite{NREL2020}, unless otherwise noted. Wind, solar PV and gas generation capital costs have been de-rated by 72\%, 51\%, and 70\% respectively to account for estimated capital cost differences for these resources between U.S. and India, as per central technology cost for 2019 from \cite{IEAWEO2020}. Solar costs assume DC to AC ratio of 1.34\cite{NREL2020}}
\begin{tabular}{l|c|c c c}
\hline
\multirow{2}{*}{\textbf{Resource \& Units}} & \multirow{2}{*}{\textbf{Scenario}} & %
    \multicolumn{3}{c}{\textbf{Capital Costs}}\\
 &  & \textbf{2030} & \textbf{2040} & \textbf{2050}\\
\hline
PV (\$/kW AC) & Reference & \multicolumn{1}{r}{558} & \multicolumn{1}{r}{407} & \multicolumn{1}{r}{369} \\
Wind (\$/kW AC) & Reference & \multicolumn{1}{r}{995} & \multicolumn{1}{r}{843} & \multicolumn{1}{r}{754} \\
    \multirow{2}{*}{Li-ion storage - energy (\$/kWh)} & Reference & \multicolumn{1}{r}{206} & \multicolumn{1}{r}{168} & \multicolumn{1}{r}{136} \\
          & Low cost & \multicolumn{1}{r}{160} & \multicolumn{1}{r}{105} & \multicolumn{1}{r}{82} \\
    \multirow{2}{*}{Li-ion storage - power (\$/kW AC)} & Reference & \multicolumn{1}{r}{179} & \multicolumn{1}{r}{137} & \multicolumn{1}{r}{119} \\
          & Low cost & \multicolumn{1}{r}{139} & \multicolumn{1}{r}{92} & \multicolumn{1}{r}{72} \\
    CCGT (\$/kW ) & Reference & \multicolumn{1}{r}{706} & \multicolumn{1}{r}{675} & \multicolumn{1}{r}{655} \\
    OCGT (\$/kW) & Reference & \multicolumn{1}{r}{647} & \multicolumn{1}{r}{616} & \multicolumn{1}{r}{598} \\
    Nuclear \cite{CEANEP2018} (\$/kW) & Reference & \multicolumn{3}{c}{2,800} \\
    Coal \cite{IEAWEO2020} (\$/kW) & Reference & \multicolumn{3}{c}{1,200} \\
    Biomass \cite{CEANEP2018} (\$/kW) & Reference & \multicolumn{3}{c}{864} \\
    Inter-regional transmission (\$/MW-km)\cite{NREL2020} & Reference & \multicolumn{3}{c}{312} \\
    \hline
% etc. ...
\end{tabular}
  \label{table:capcost}%
\end{table}%

\end{document}

{\centering
\textbf{\emph{\Huge{Supplementary Information}}}
}

\listoffigures
\listoftables

\vspace{20cm}

\begin{figure}[!ht]
\centering
 \includegraphics[width=15cm,height=15cm, keepaspectratio]{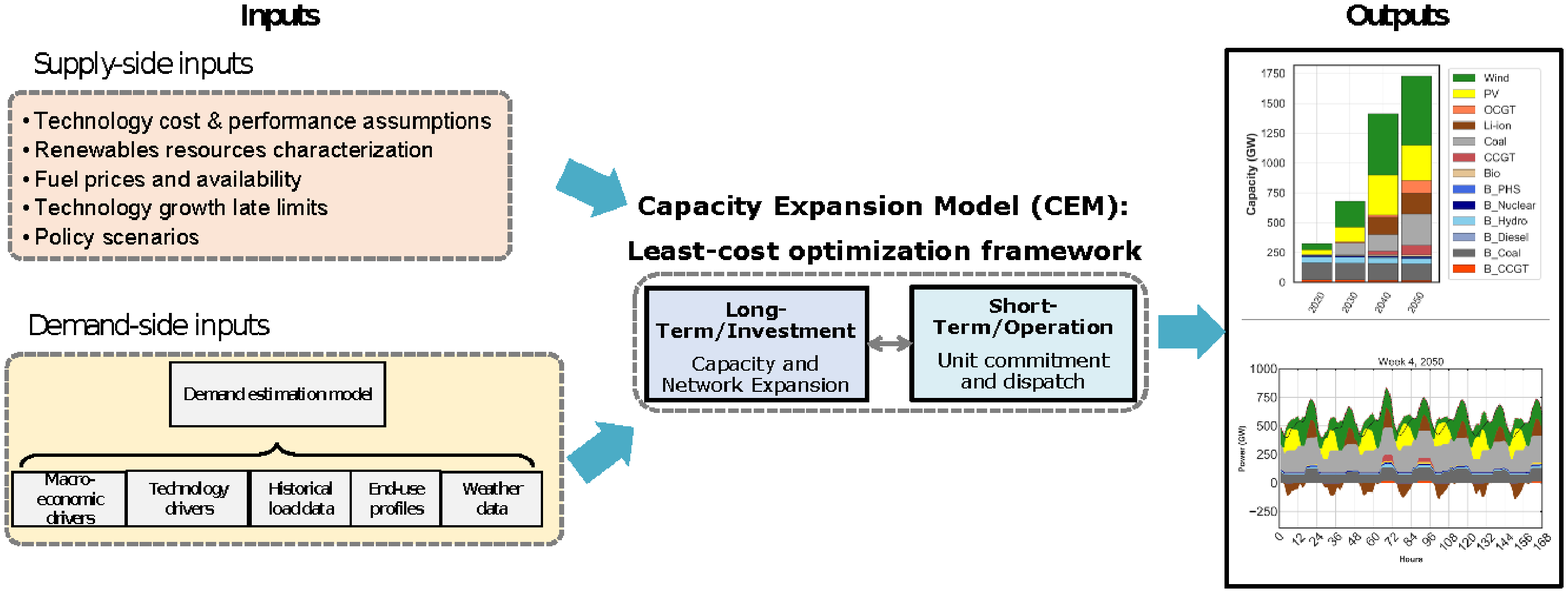}
  \caption{Methodologfy schematic. Demand forecasting model is documented in \cite{Barbar2021} while GenX model documentation and description is available here \cite{gitgenx}}
   \label{fig:schematic}
\end{figure}

\begin{figure}[!ht]
\centering
 \includegraphics[width=15cm,height=15cm, keepaspectratio]{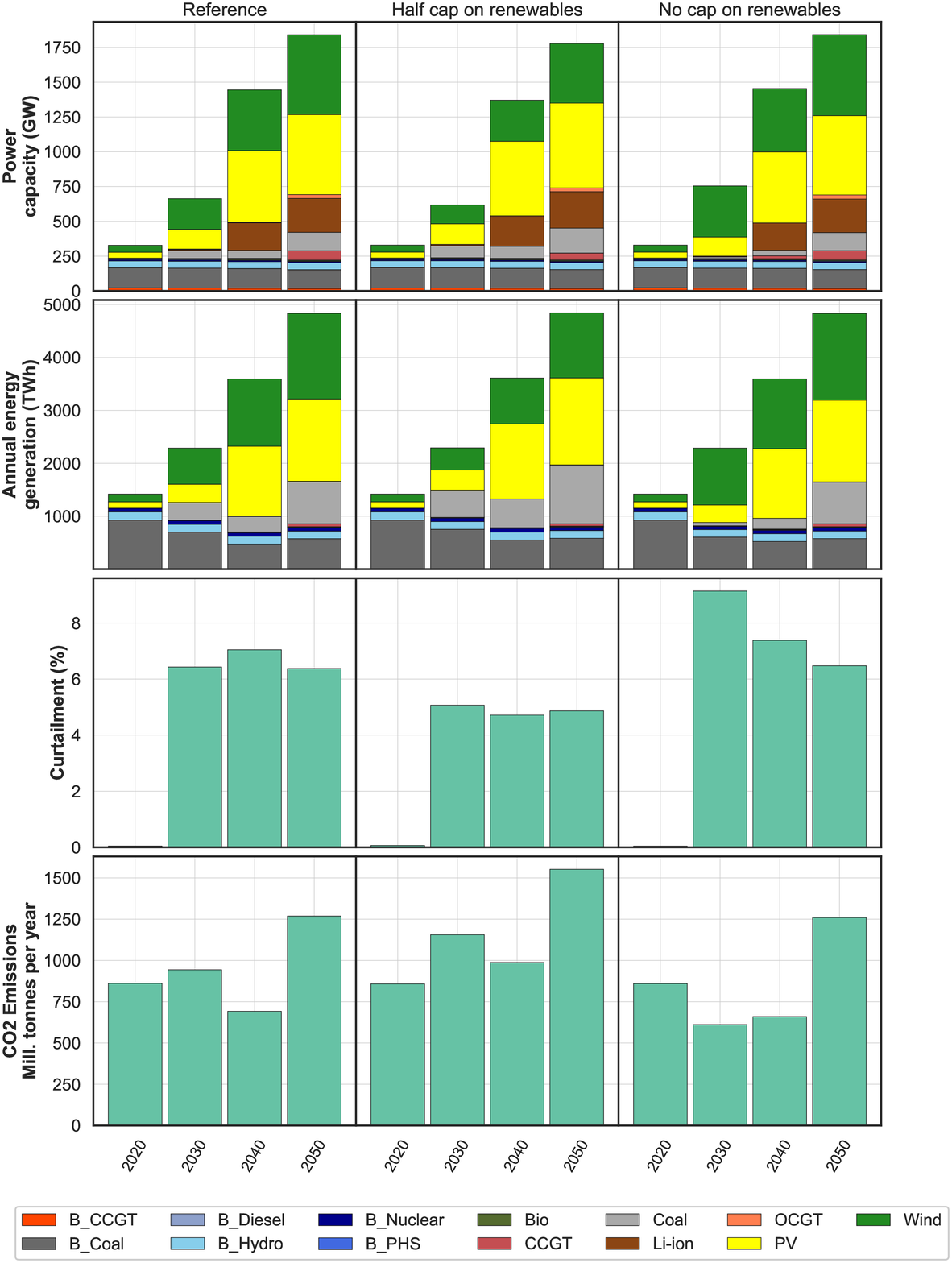}
  \caption{Reference case model outcomes with alternative assumptions about decadal renewables installation limits. Reference = decadal installation limits as shown in Table \ref{table:caplimit}. Half cap = decadal installation limits are 0.5 of the reference values. No cap = no decadal installation limits.}
   \label{fig:10_SI}
\end{figure}

\begin{figure}[!ht]
\centering
 \includegraphics[width=15cm,height=15cm, keepaspectratio]{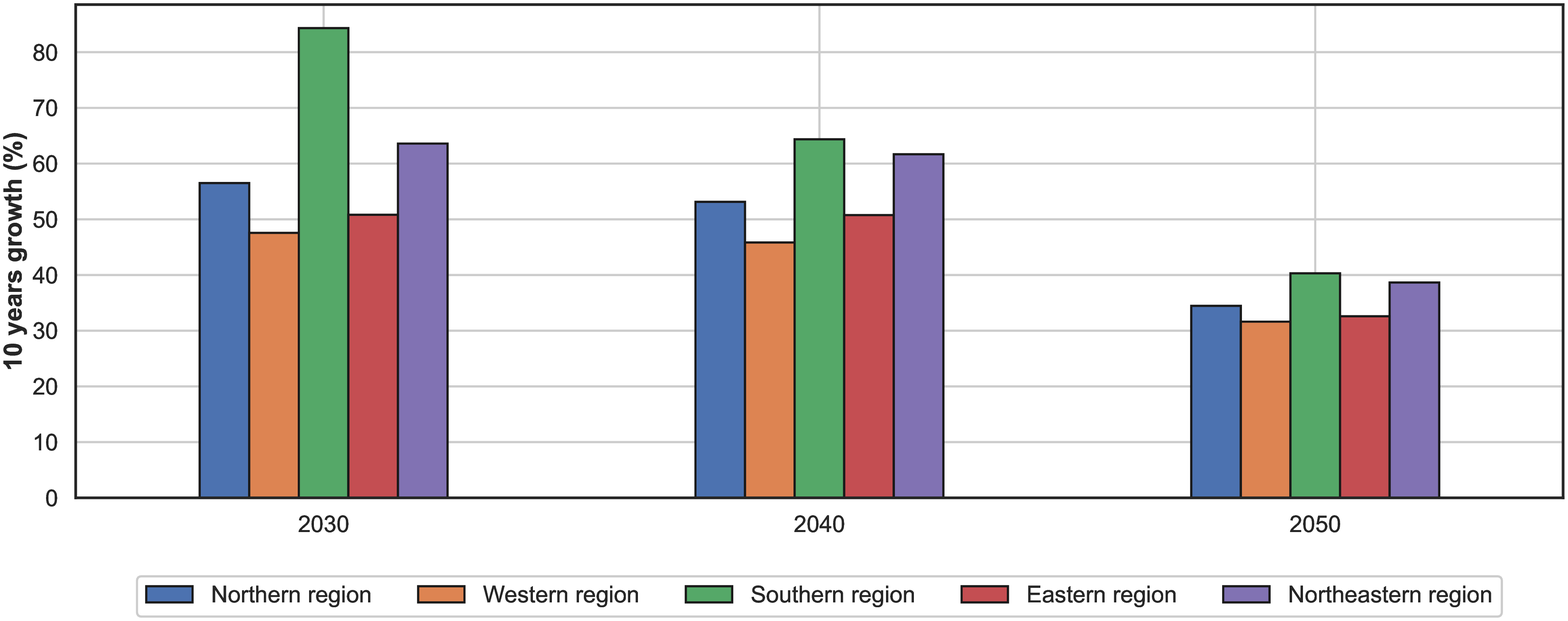}
  \caption{Regional demand growth projections for India as per demand forecasting model results \cite{Barbar2021}}
   \label{fig:17_SI}
\end{figure}

\begin{figure}[!ht]
\centering
 \includegraphics[width=15cm,height=15cm, keepaspectratio]{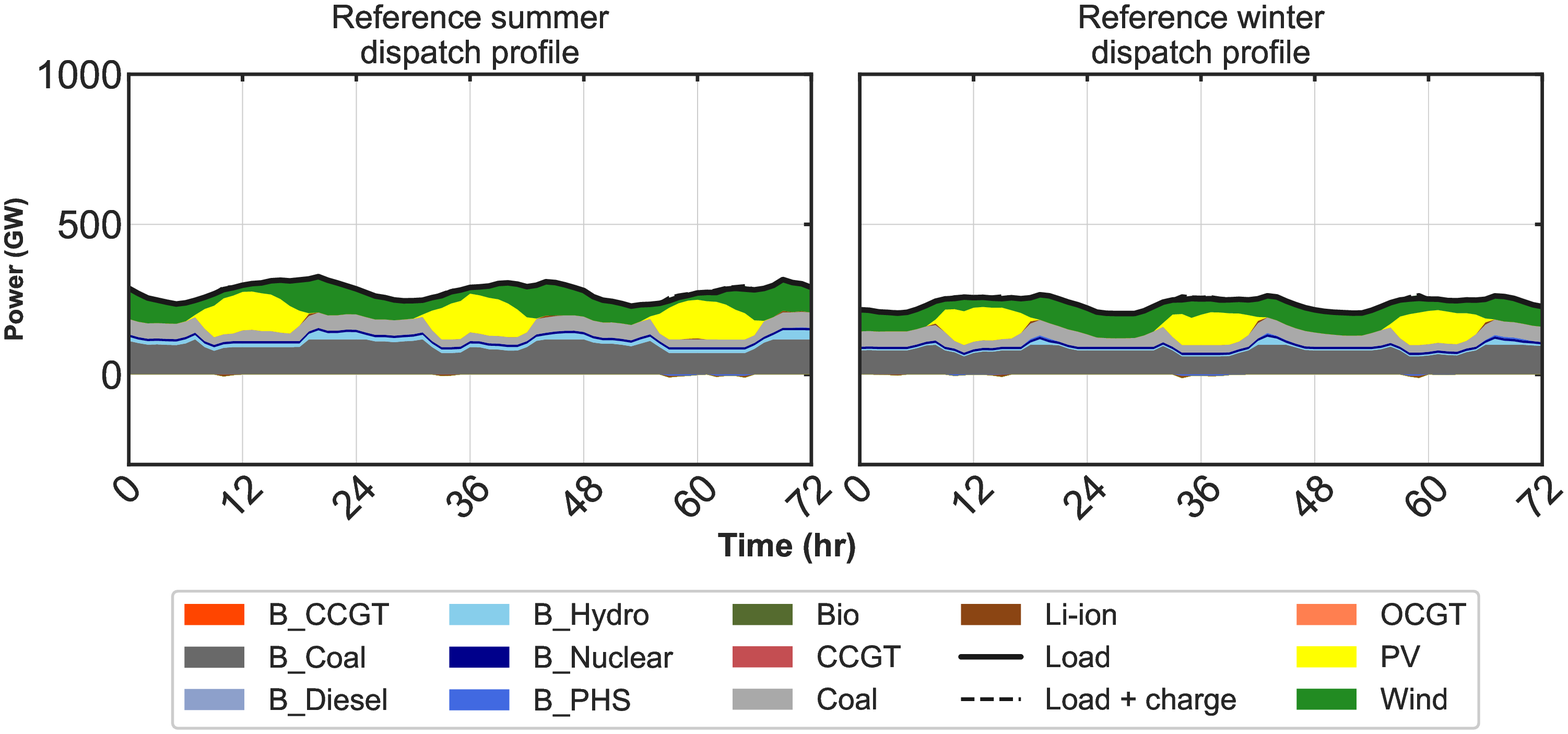}
  \caption{Hourly generation dispatch for reference winter and summer load profiles for 2030. Technology names and their respective abbreviations in Supplementary Table \ref{table:abr}}
   \label{fig:13_4_SI}
\end{figure}

\begin{figure}[!ht]
\centering
 \includegraphics[width=15cm,height=15cm, keepaspectratio]{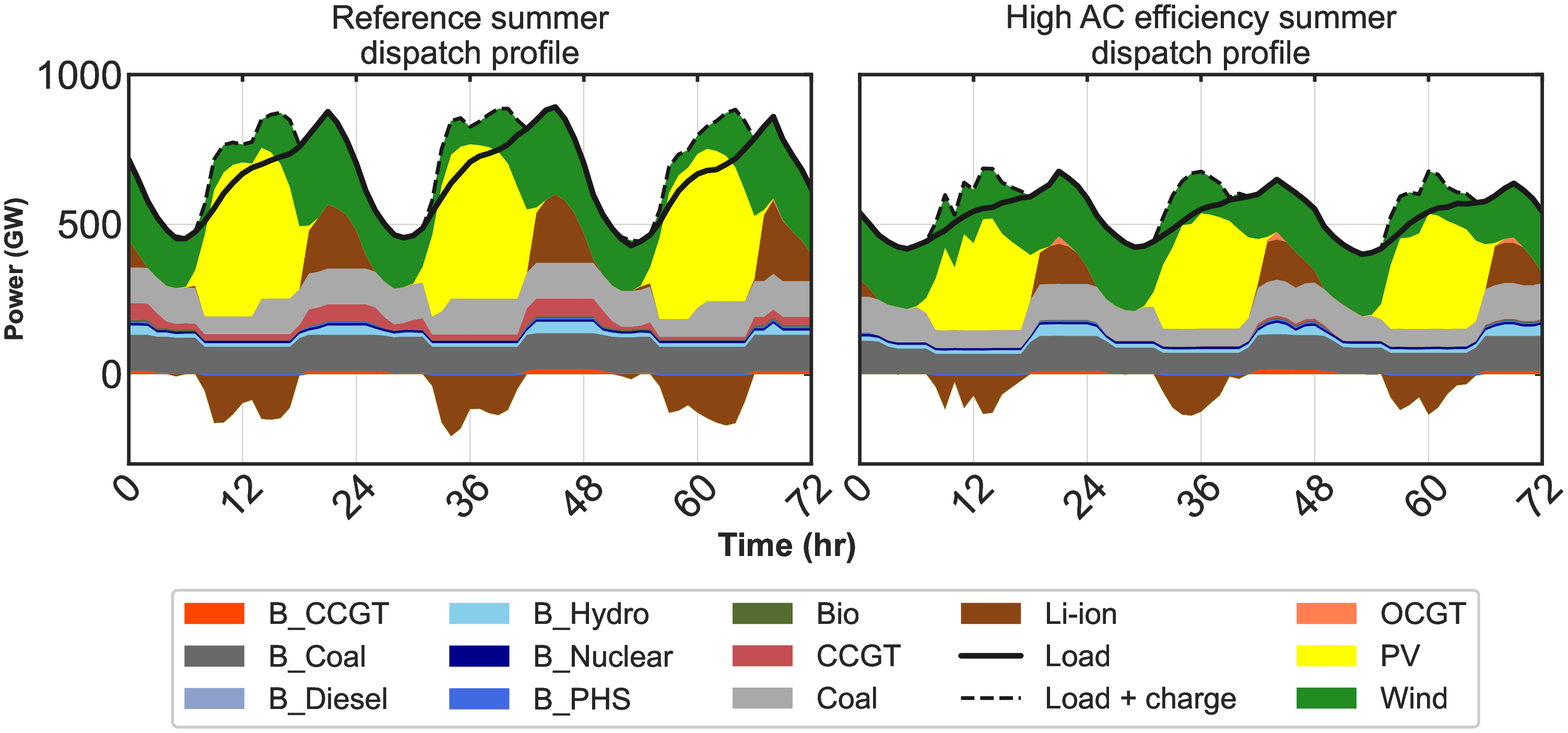}
  \caption{Hourly generation dispatch for summer reference and high AC efficiency load profiles for 2050. Technology names and their respective abbreviations in Supplementary Table \ref{table:abr}}
   \label{fig:13_5_SI}
\end{figure}

\begin{figure}[!ht]
\centering
 \includegraphics[width=15cm,height=15cm, keepaspectratio]{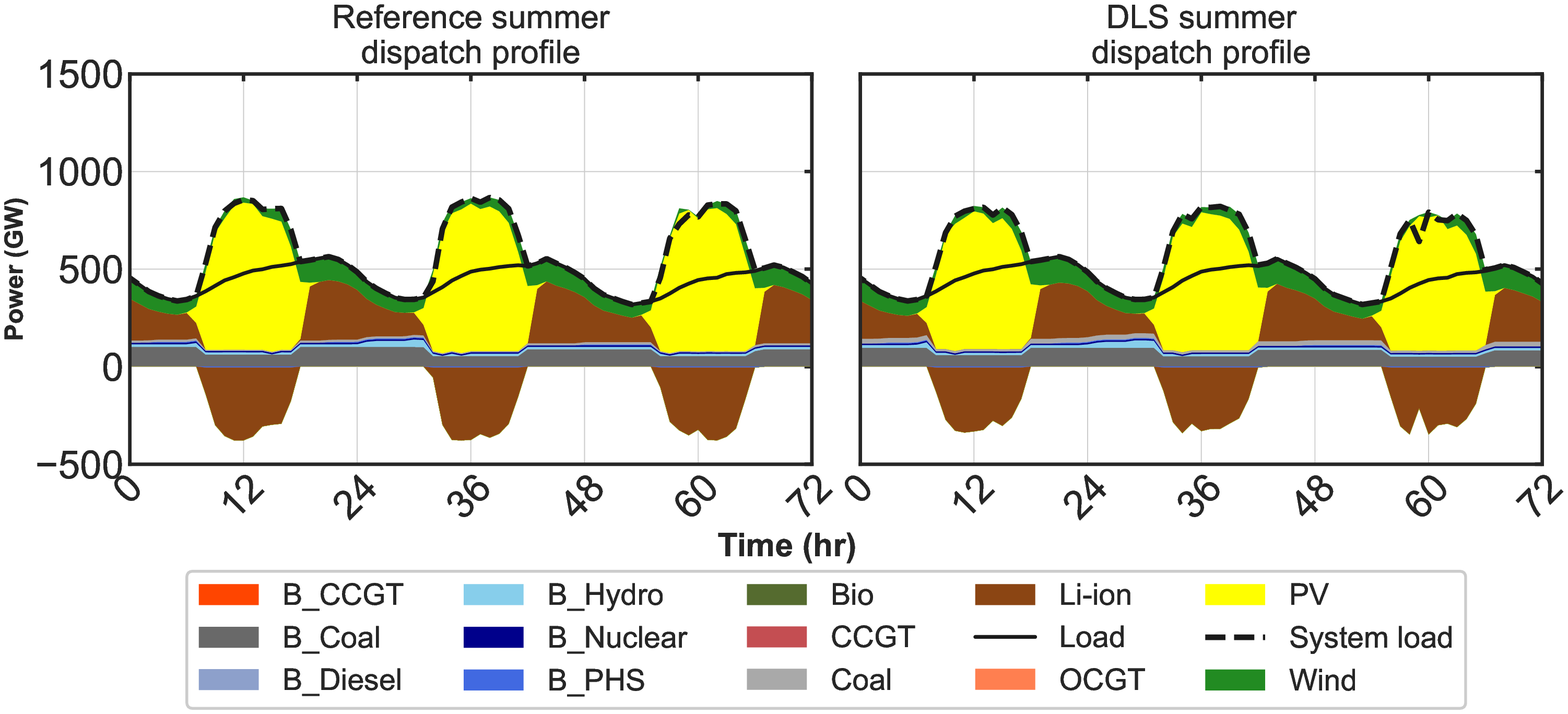}
  \caption{Hourly generation dispatch for summer reference and DLS load profiles for 2040. Technology names and their respective abbreviations in Supplementary Table \ref{table:abr}}
   \label{fig:13_6_SI}
\end{figure}
% Add a dispatch chart for High AC efficiency case.

\begin{figure}[!ht]
\centering
 \includegraphics[width=15cm,height=15cm, keepaspectratio]{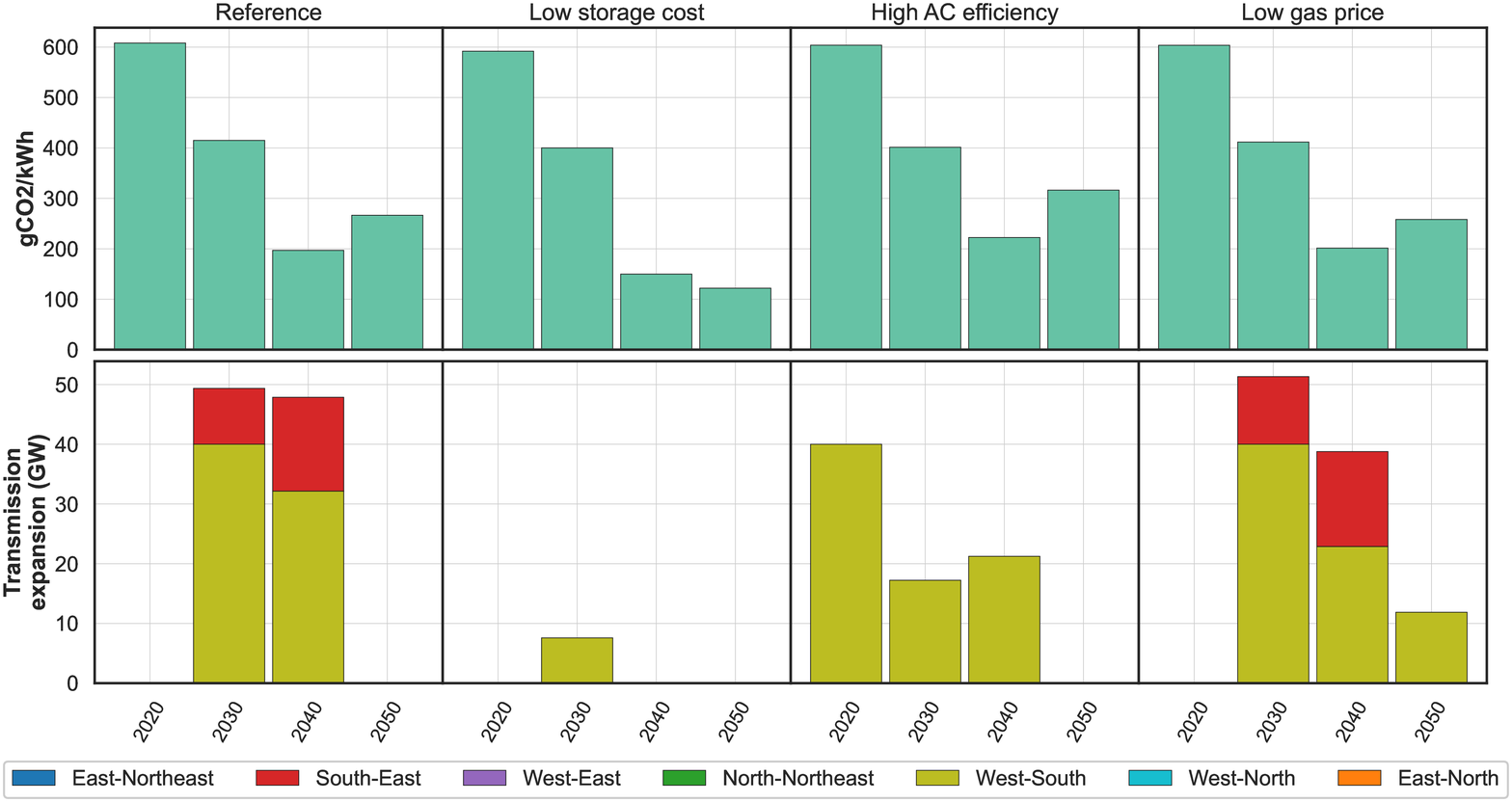}
  \caption{Emissions intensity and Transmission expansion outcomes for modeled cases considered in the main text}.
   \label{fig:9_SI}
\end{figure}

\begin{figure}[!ht]
\centering
 \includegraphics[width=15cm,height=15cm, keepaspectratio]{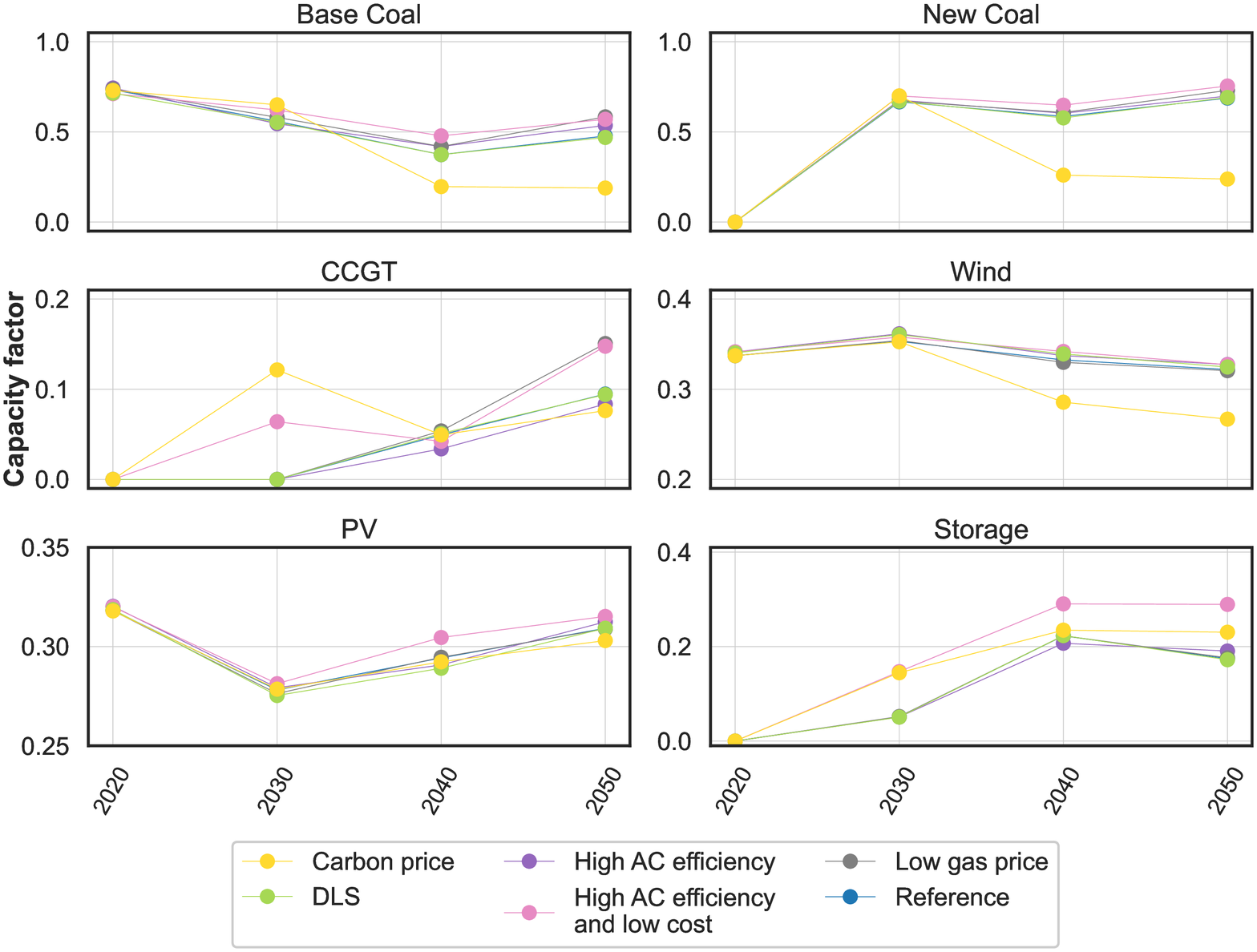}
  \caption{Technology capacity factors across the scenarios over different modeling periods}
   \label{fig:11_SI}
\end{figure}

%\begin{figure}[!ht]
%\centering
% \includegraphics[width=15cm,height=15cm, keepaspectratio]{Figure_13_2_SI.eps}
%  \caption{Reference case cumulative distribution function of hourly load variation}
%   \label{fig:13_2_SI}
%\end{figure}

\begin{figure}[!ht]
\centering
 \includegraphics[width=15cm,height=15cm, keepaspectratio]{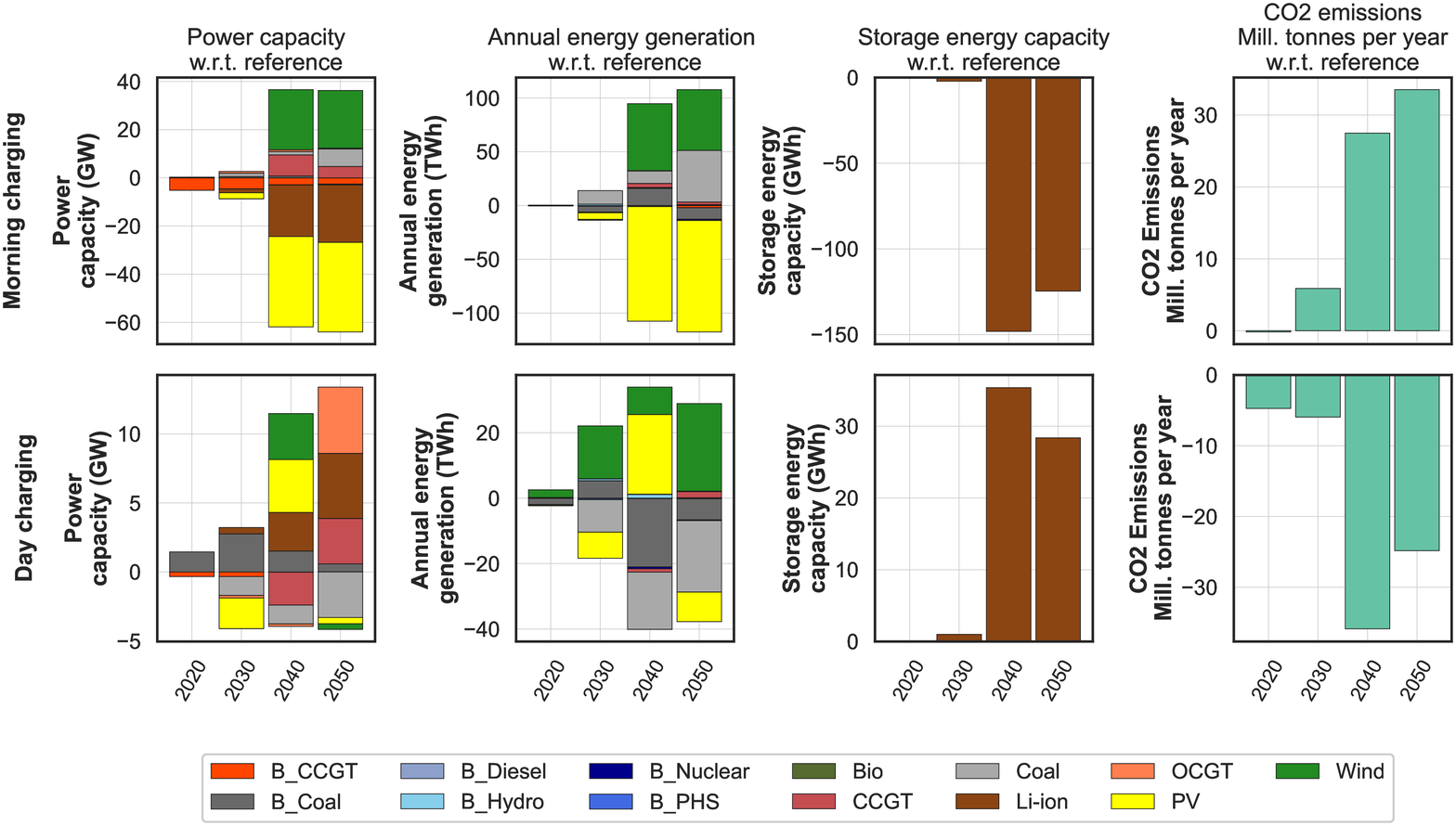}
  \caption{Impact of Morning (top) and day (bottom) electric vehicle (EV) charging schemes relative to evening EV charging scheme under the reference case}
   \label{fig:ev}
\end{figure}

%\begin{figure}[!ht]
%\centering
% \includegraphics[width=15cm,height=15cm, keepaspectratio]{Figure_4_SI.eps}
%  \caption{Model outcomes for sensitivity scenario, defined by low battery storage capital cost, high AC efficiency and low gas price (1$^{st}$ column) as well as impact of carbon price with and without sensitivity scenario assumptions. Model outcomes include Installed capacity (1$^{st}$ row), generation (2$^{nd}$ row), energy storage capacity (3$^{rd}$ row) and annual CO$_2$ emissions (4$^{th}$ row).}
%   \label{fig:4_SI}
%\end{figure}

\begin{figure}[!ht]
\centering
 \includegraphics[width=15cm,height=15cm, keepaspectratio]{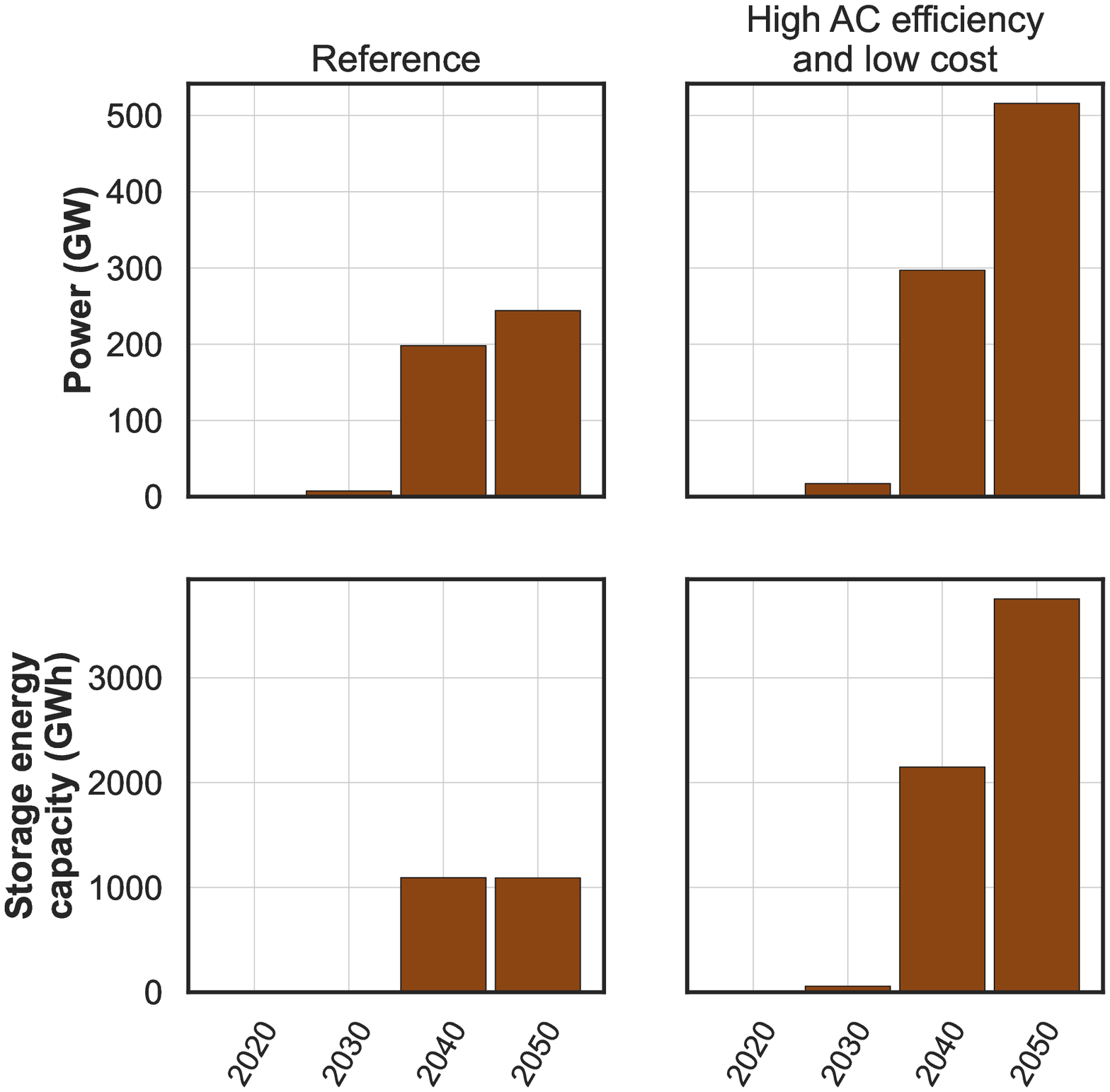}
  \caption{Storage power and energy capacity deployment trends in the reference and sensitivity cases}
   \label{fig:21_SI}
\end{figure}

\begin{landscape}
\begin{figure}[!ht]
\centering
 \includegraphics[width=20cm,height=20cm, keepaspectratio]{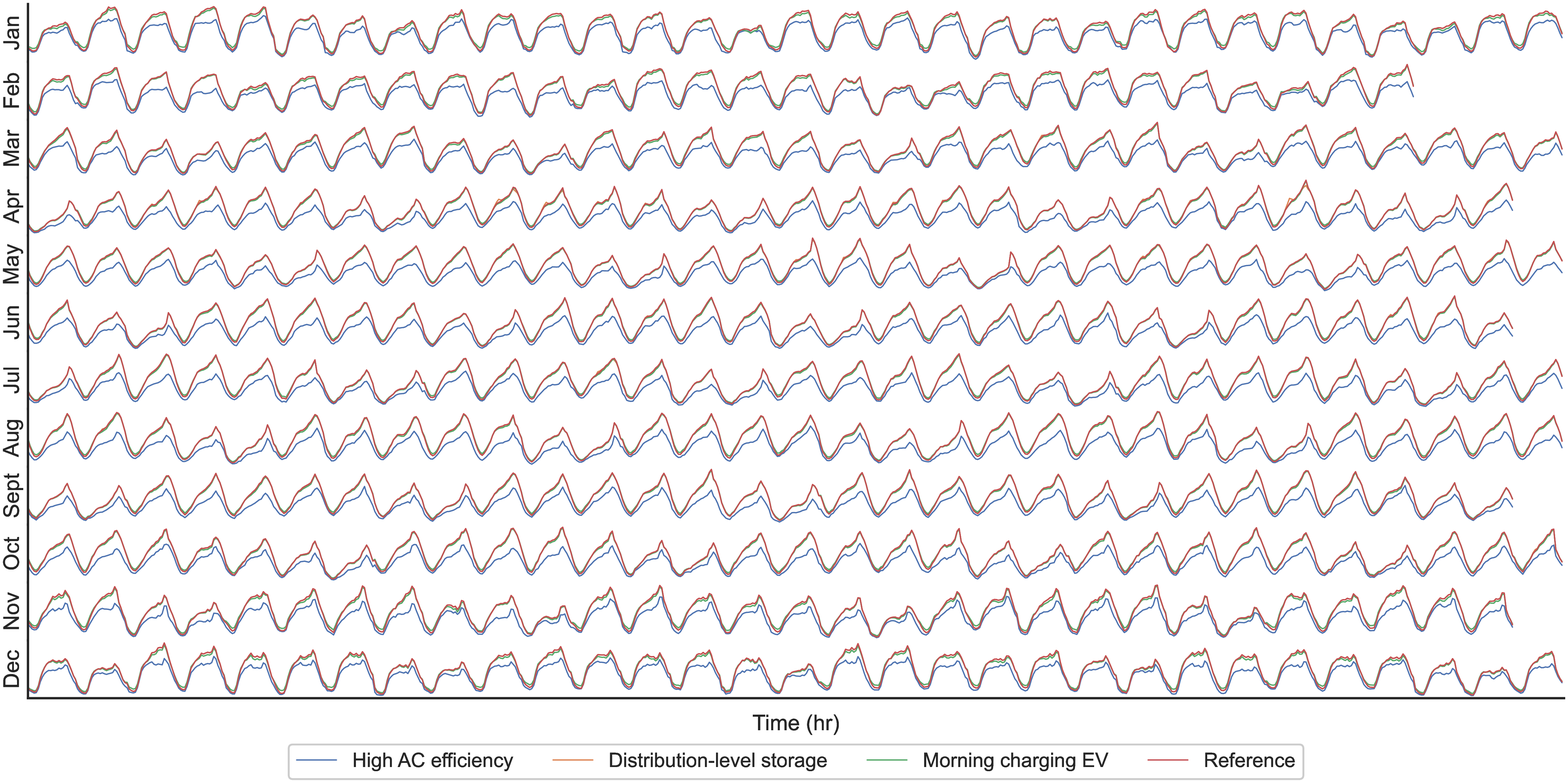}
  \caption{Hourly load profile by month in 2040 across various demand scenarios considered here}
   \label{fig:14_SI}
\end{figure}
\end{landscape}

\begin{figure}[!ht]
\centering
 \includegraphics[width=15cm,height=15cm, keepaspectratio]{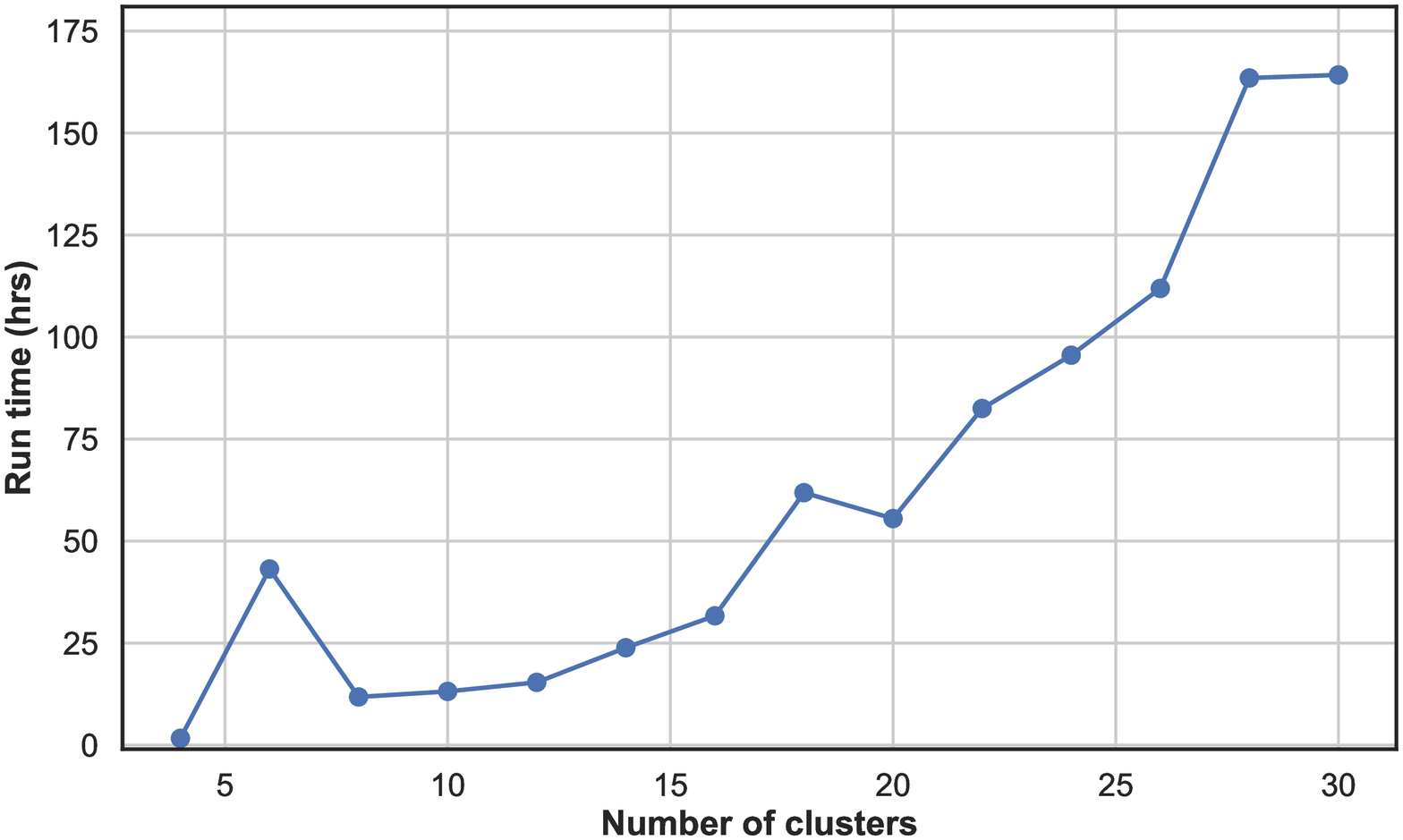}
  \caption{Indicative GenX capacity expansion optimization model run time with respect to number of clustered weeks. Outputs based on reference case.}
   \label{fig:15_SI}
\end{figure}

\begin{figure}[!ht]
\centering
 \includegraphics[width=15cm,height=15cm, keepaspectratio]{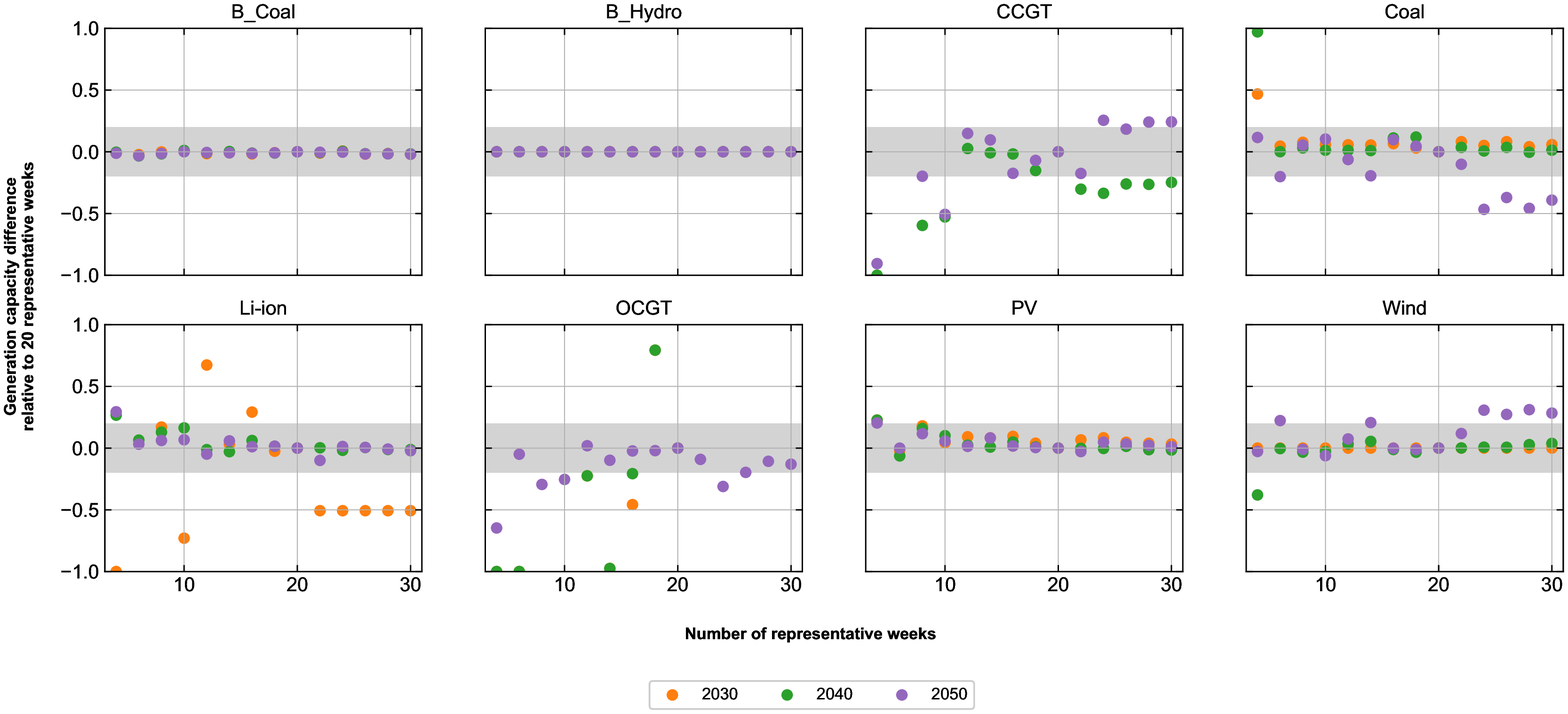}
  \caption{Generation capacity difference relative to 20 representative weeks for the reference case.}
   \label{fig:20_SI}
\end{figure}

%\begin{figure}[!ht]
%\centering
% \includegraphics[width=15cm,height=15cm, keepaspectratio]{Figure_16_SI.eps}
%  \caption{GenX capacity expansion optimization model optimality gap with respect to cumulative run time}
%   \label{fig:16_SI}
%\end{figure}

\begin{figure}[!ht]
\centering
 \includegraphics[width=15cm,height=15cm, keepaspectratio]{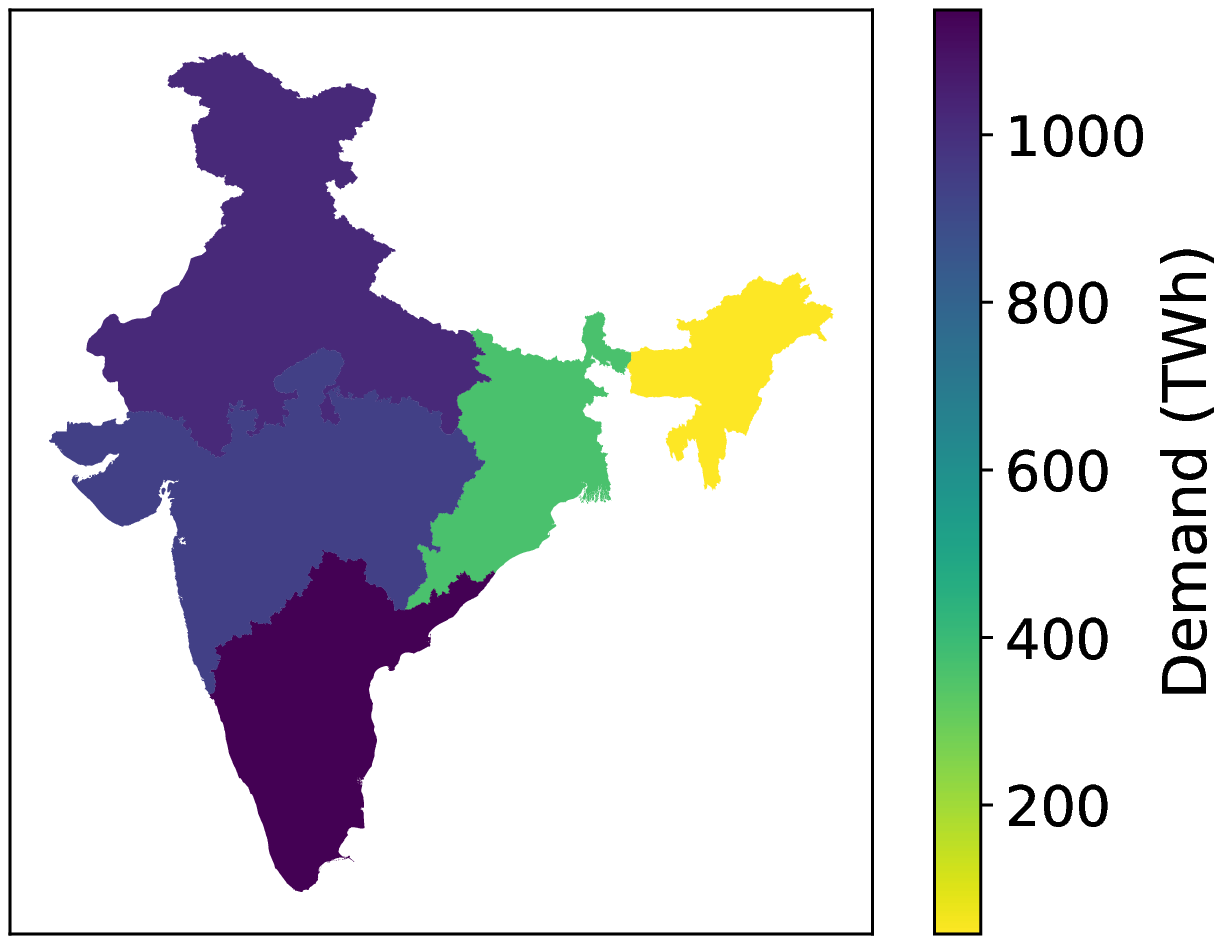}
  \caption{Regional demand under the reference case in 2040}
   \label{fig:18_SI}
\end{figure}

\begin{figure}[!ht]
\centering
 \includegraphics[width=15cm,height=15cm, keepaspectratio]{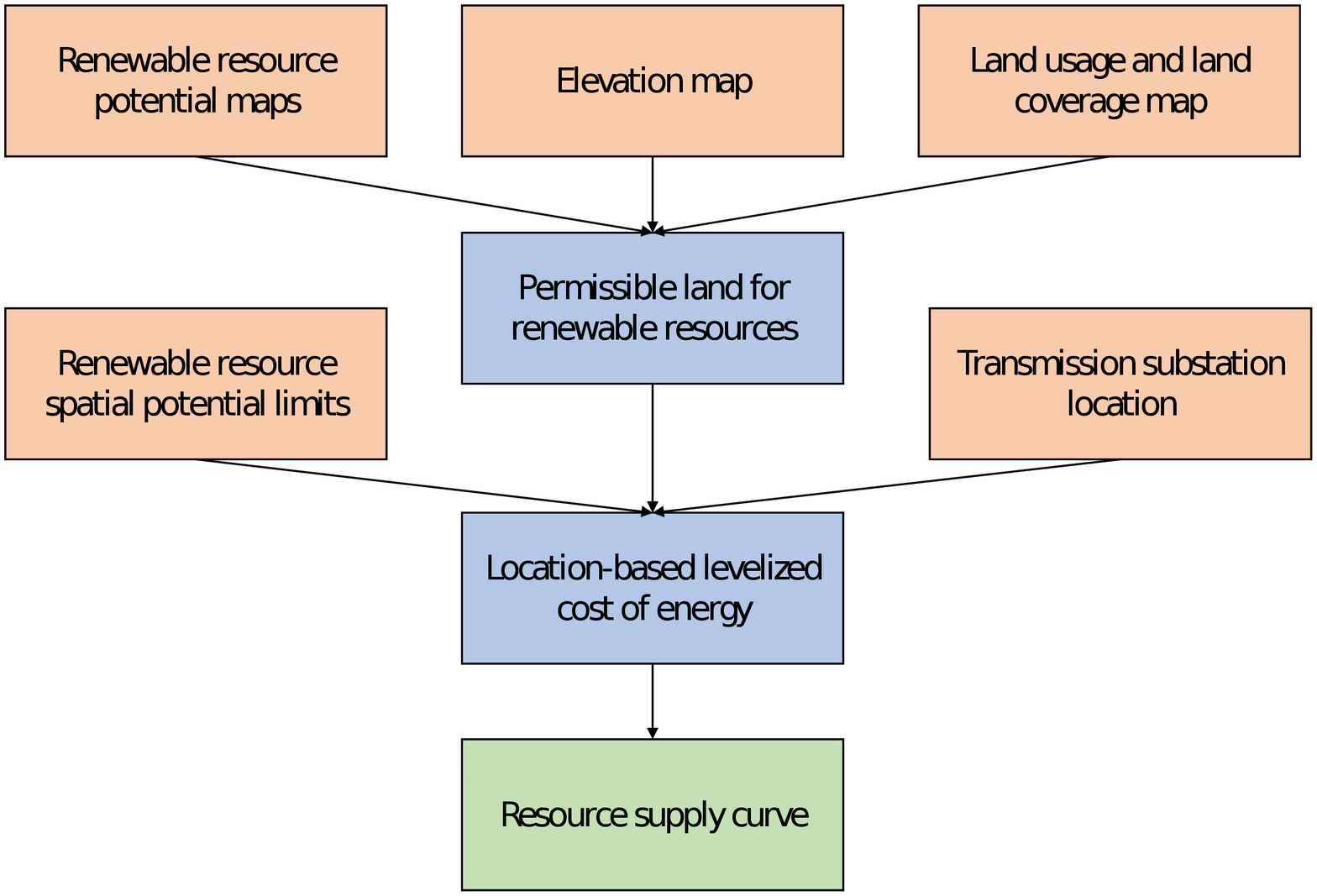}
  \caption{Renewable resources supply curve calculation flowchart}
   \label{fig:flowchart_SI}
\end{figure}

\begin{figure}[!ht]
\centering
 \includegraphics[width=15cm,height=15cm, keepaspectratio]{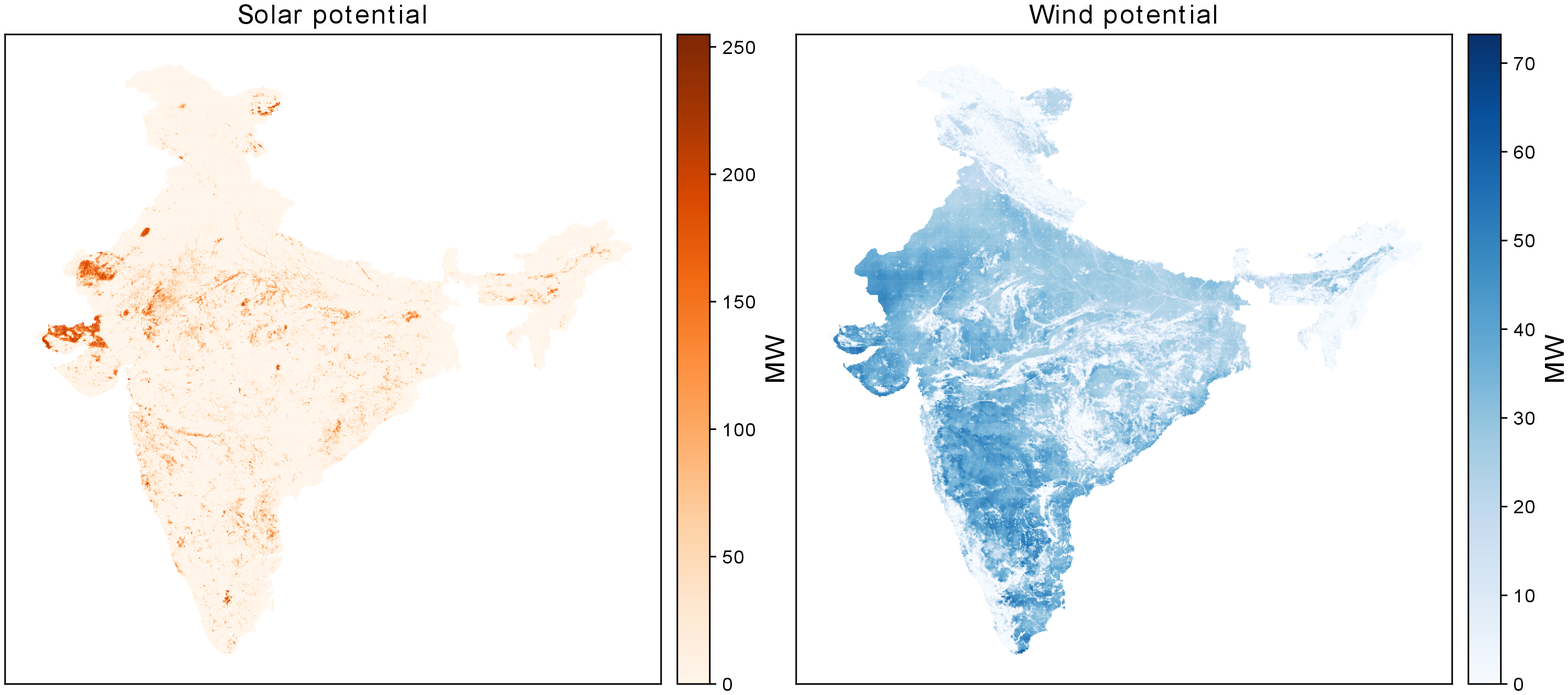}
  \caption{Deployable solar and wind resources potential maps}
   \label{fig:19_SI}
\end{figure}

\begin{figure}[!ht]
\centering
 \includegraphics[width=15cm,height=15cm, keepaspectratio]{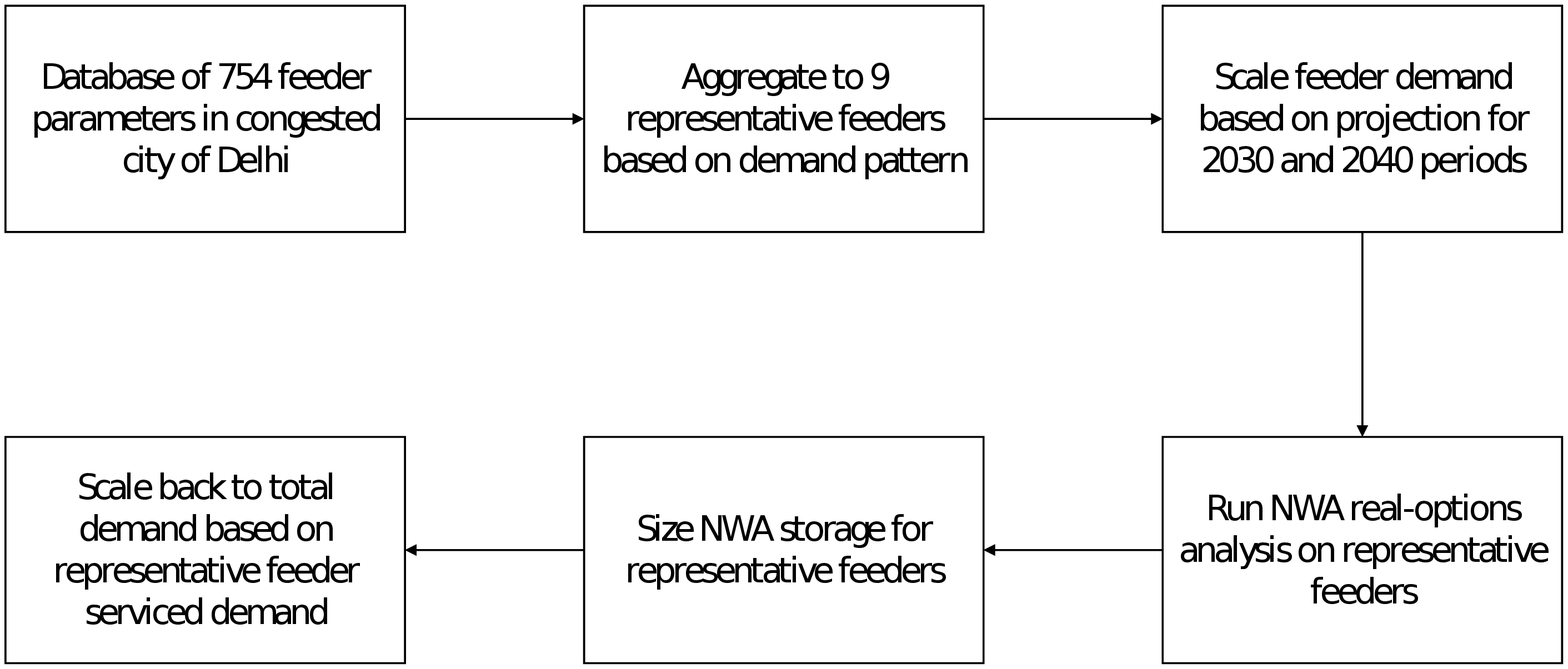}
  \caption{DLS method flowchart}
   \label{fig:dls-flow}
\end{figure}

%\begin{figure}[!ht]
%\centering
% \includegraphics[width=15cm,height=15cm, keepaspectratio]{Figure_12_SI.eps}
%  \caption{Reference case normalized national load duration curve}
%   \label{fig:12_SI}
%\end{figure}

\begin{figure}[!ht]
\centering
 \includegraphics[width=15cm,height=15cm, keepaspectratio]{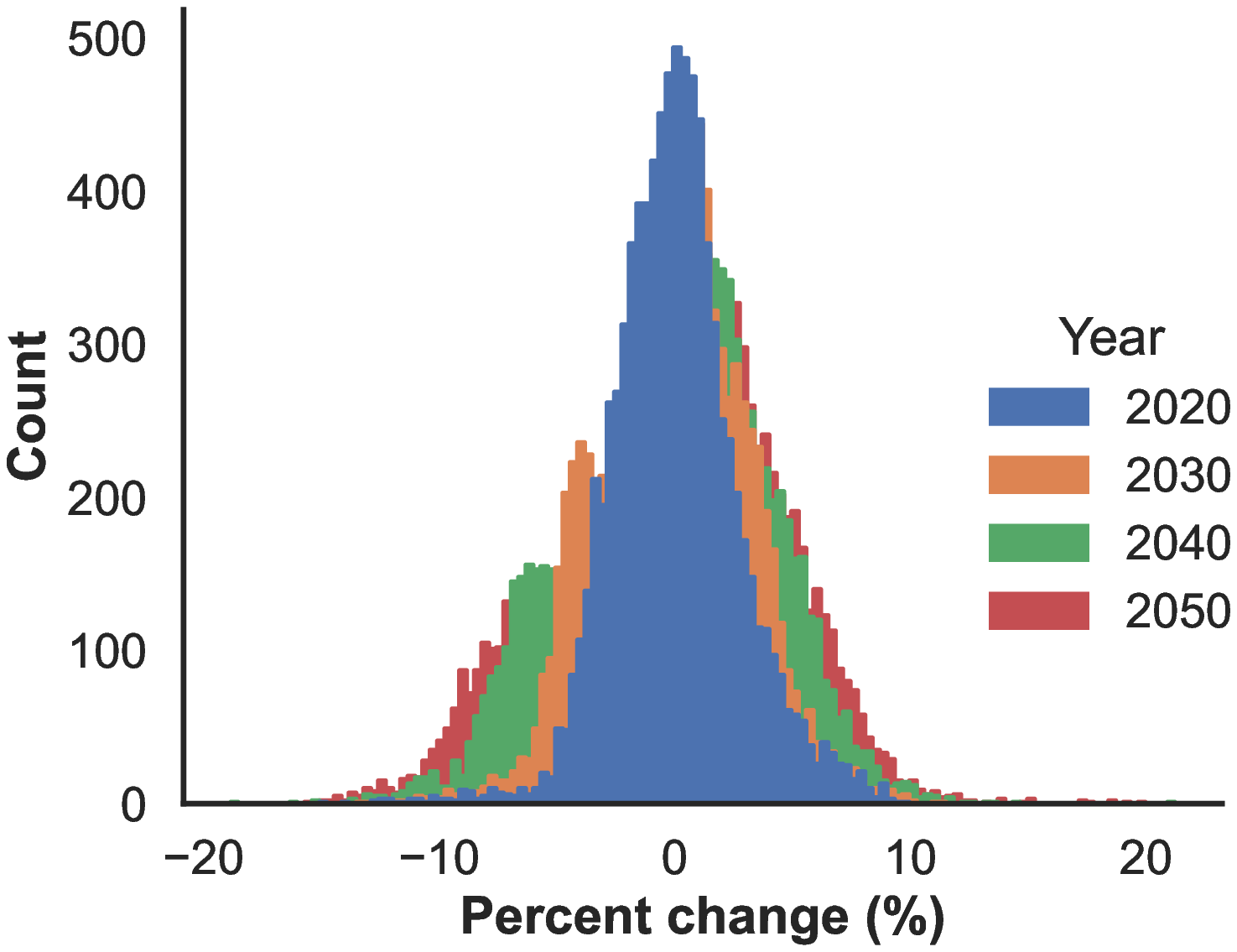}
  \caption{Reference case frequency of hourly load variation}
   \label{fig:13_1_SI}
\end{figure}

%\begin{figure}[!ht]
%\centering
% \includegraphics[width=15cm,height=15cm, keepaspectratio]{options-framework.eps}
%  \caption{Flexible valuation framework steps flowchart}
%   \label{fig:options-framework}
%\end{figure}

%\begin{figure}[!ht]
%\centering
% \includegraphics[width=15cm,height=15cm, keepaspectratio]{options-flowchart.eps}
%  \caption{Real options decision tree}
%   \label{fig:options-flowchart}
%\end{figure}

%%%%%%%%%%%%%%%%%%%%%%%%%%%%%%%%%%%%%%%%%%%%%%%%%%%%%%%%%%%%%%%

\begin{table}[!ht]
\centering
\caption{Carbon price scenarios}
\begin{tabularx}{5cm}{l|CC}
\hline
\textbf{Year} & \textbf{Carbon price} \\
\hline
2020 & 0 \\
2030 & 20 \\
2040 & 33 \\
2050 & 53\\
\hline
\end{tabularx}
\label{table:co2}
\end{table}

\begin{table}[!ht]
  \centering
  \caption{Capital cost assumptions for various resources. All costs in 2018 dollars and sourced from NREL annual technology baseline 2020 \cite{NREL2020}, unless otherwise noted. Wind, solar PV and gas generation capital costs have been de-rated by 72\%, 51\%, and 70\% respectively to account for estimated capital cost differences for these resources between U.S. and India, as per central technology cost for 2019 from \cite{IEAWEO2020}. Solar costs assume DC to AC ratio of 1.34\cite{NREL2020}}
\begin{tabular}{l|c|c c c}
\hline
\multirow{2}{*}{\textbf{Resource \& Units}} & \multirow{2}{*}{\textbf{Scenario}} & %
    \multicolumn{3}{c}{\textbf{Capital Costs}}\\
 &  & \textbf{2030} & \textbf{2040} & \textbf{2050}\\
\hline
PV (\$/kW AC) & Reference & \multicolumn{1}{r}{558} & \multicolumn{1}{r}{407} & \multicolumn{1}{r}{369} \\
Wind (\$/kW AC) & Reference & \multicolumn{1}{r}{995} & \multicolumn{1}{r}{843} & \multicolumn{1}{r}{754} \\
    \multirow{2}{*}{Li-ion storage - energy (\$/kWh)} & Reference & \multicolumn{1}{r}{206} & \multicolumn{1}{r}{168} & \multicolumn{1}{r}{136} \\
          & Low cost & \multicolumn{1}{r}{160} & \multicolumn{1}{r}{105} & \multicolumn{1}{r}{82} \\
    \multirow{2}{*}{Li-ion storage - power (\$/kW AC)} & Reference & \multicolumn{1}{r}{179} & \multicolumn{1}{r}{137} & \multicolumn{1}{r}{119} \\
          & Low cost & \multicolumn{1}{r}{139} & \multicolumn{1}{r}{92} & \multicolumn{1}{r}{72} \\
    CCGT (\$/kW ) & Reference & \multicolumn{1}{r}{706} & \multicolumn{1}{r}{675} & \multicolumn{1}{r}{655} \\
    OCGT (\$/kW) & Reference & \multicolumn{1}{r}{647} & \multicolumn{1}{r}{616} & \multicolumn{1}{r}{598} \\
    Nuclear \cite{CEANEP2018} (\$/kW) & Reference & \multicolumn{3}{c}{2,800} \\
    Coal \cite{IEAWEO2020} (\$/kW) & Reference & \multicolumn{3}{c}{1,200} \\
    Biomass \cite{CEANEP2018} (\$/kW) & Reference & \multicolumn{3}{c}{864} \\
    Inter-regional transmission (\$/MW-km)\cite{NREL2020} & Reference & \multicolumn{3}{c}{312} \\
    \hline
% etc. ...
\end{tabular}
  \label{table:capcost}%
\end{table}%

\begin{table}[!ht]
\centering
\caption{Modeled technology abbreviations}
\begin{tabularx}{15cm}{l|C}
\hline
\multicolumn{1}{C|}{\textbf{Technology}} & \textbf{Abbreviation}  \\
\hline
Existing combined cycle gas turbine generation & B\_CCGT  \\
Existing coal generation & B\_Coal  \\
Existing diesel generation & B\_Diesel  \\
Existing hydro generation & B\_Hydro  \\
Existing nuclear generation & B\_Nuclear  \\
Existing pumped hydro storage generation & B\_PHS  \\
New biomass generation & Bio  \\
New combined cycle gas turbine generation & CCGT  \\
New coal generation & Coal \\
New Lithium ion battery storage power & Li-ion \\
New open cycle gas turbine generation & OCGT  \\
New solar generation & PV \\
New wind generation & Wind  \\
Network expansion & NetworkExp \\
\hline
\end{tabularx}
\label{table:abr}
\end{table}

\begin{table}[!ht]
  \centering
  \caption{System assumptions}
    \begin{tabularx}{12cm}{l|C|C}
    \hline
    \multicolumn{1}{l|}{\textbf{Parameter}} & \multicolumn{2}{c}{\textbf{Value}} \\
    \hline
    Discount rate  & \multicolumn{2}{c}{9\%} \\
    
    Value of lost load (\$/MWh) & \multicolumn{2}{c}{20,000} \\
    \hline
    \hline
    \multirow{2}{*}{Fuels assumptions} & \multicolumn{1}{C|}{Cost (\$/MMBtu)} & \multicolumn{1}{C}{Emissions intensity (tonnes CO$_2$/MMBtu)} \\
    \hline
    Uranium& 1   & 0.000 \\
    Coal & 3   & 0.096 \\
    Natural gas (Reference / low) & 11 / 8  & 0.052 \\
    Diesel & 18  & 0.073 \\
    Biomass & 3.7 & 0.000 \\
    \hline
    \end{tabularx}%
  \label{table:sys}%
\end{table}%

\begin{table}[!ht]
\centering
\caption{Storage duration comparison across scenarios and modeled periods}
\begin{tabular}{l|ccccc}
\hline
 & \textbf{Reference} & \textbf{Low storage cost} & \textbf{High AC efficiency} & \textbf{Low gas price} & \textbf{Sensitivity} \\
 \hline
2040 & 4.5 & 5.4 & 4.4 & 4.6 & 6.8 \\
2050 & 3.7 & 5.5 & 3.7 & 3.6 & 6.2 \\
\hline
\end{tabular}
\label{table:bat_duration}
\end{table}

\begin{landscape}

\begin{table}[!ht]
\centering
\caption{Regional thermal power existing capacity and parameters \cite{NREL2020, ATB2020, osti_1563140, DESHMUKH2019947, OSM2020, WB2020, LULC16}}
\begin{tabularx}{22cm}{l|llCccccCCCC}
\hline
 & & & & \multicolumn{3}{c}{\textbf{Minimum Retirement   (MW)}} &  &  \\
 \textbf{Resource} & \textbf{Fuel} & \textbf{Region} & \textbf{2020 Capacity (MW)} & \textbf{2030} & \textbf{2040} & \textbf{2050} & \textbf{VOM (\$/MWh)} & \textbf{FOM (\$/kW/y)}  & \textbf{Heat rate (MMBtu per MWh)} & \textbf{Average plant size (MW)} & \textbf{Maximum Capacity (MW)} \\
 \hline
 & & North & 41,220 & 3,919 & 11,755 & 13,588 & 1.0 & 55.1 &  10.0 & 400 & \\
 & & West & 87,431 & 5,812 & 28,890 & 37,398 & 0.9 & 55.1 & 9.1 & 400 & \\
 & & South & 40,965 & 5,150 & 10,868 & 16,368 & 1.1 & 55.1 &  10.6 & 450& \\
 & & East & 39,080 & 4,010 & 5,295 & 5,895 & 0.9 & 55.0 &  10.5 & 440& \\

\multirow{-5}{*}{\textbf{Coal}} & \multirow{-5}{*}{Coal} & Northeast & 750 & 0 & 0 & 250 & 1.0 & 55.0 &   9.8 & 230& \\
 \hline
 & & North & 5,752 & 179 & 685 & 910 & 1.2 & 9.4 & 7.8 & 480& \\
 & & West & 10,239 & 870 & 2,686 & 3,022 & 1.5 & 12.0 &  6.9 & 410& \\
 & & South & 6,505 & 1,147 & 1,948 & 2,115 & 1.4 & 11.0 &   6.2 & 470& \\

\multirow{-4}{*}{\textbf{CCGT}} & \multirow{-4}{*}{Natural gas} & Northeast & 1,306 & 19 & 351 & 390 & 1.9 & 10.8 &  7.7 & 140&  \\
 \hline
 & & North & 1,720 &  &  &  &  &  &  &   \\
 & & West & 3,240 &  &  &  &  &  &  &   \\

\multirow{-3}{*}{\textbf{Nuclear}} & \multirow{-3}{*}{Uranium} & South & 3,820 &  &  &  &  &  &  & & \\
 \hline
 & & North & 2,431 &  &  &  &  &  &  & & 9,721  \\
 & & West & 678.75 &  &  &  &  &  &  & & 6,835 \\
 & & South & 2,934 &  &  &  &  &  &  & & 5,336 \\
 & & East & 463 &  &  &  &  &  &  & & 1,906 \\
\multirow{-5}{*}{\textbf{Biomass}} & \multirow{-5}{*}{Biomass} & Northeast & 0 &  &  &  &  &  &  & & 274\\
\hline
 & & South & 761.58 &  &  &  &  &  &  & &  \\
\multirow{-2}{*}{\textbf{Backup}} & \multirow{-2}{*}{Diesel} &  Northeast & 36 &  &  &  &  &  &  & & \\
\hline
\end{tabularx}
\label{table:reg_thermal}
\end{table}

\end{landscape}

\begin{landscape}

\begin{table}[!ht]
\centering
\caption{National thermal power parameters \cite{NREL2020, Rudnick2019, ATB2020, IEAWEO2020, osti_1563140, NRELGG2017, CEANEP2018, Mallapragada2020}}
\begin{tabularx}{23cm}{l|cCCCCCCCCCCCCC}
\hline
\textbf{Resource} & \textbf{Fuel} & \textbf{VOM (\$/MWh)} & \textbf{FOM (\$/kW/y)} & \textbf{Start cost (\$/MW)} & \textbf{Start fuel (MMBtu/ MW)} & \textbf{Heat rate (MMBtu/ MWh)} & \textbf{Min up time (hours)} & \textbf{Min down time (hours)} & \textbf{Ramp up} & \textbf{Ramp down} & \textbf{Min stable power} & \textbf{Max power} & \textbf{Average plant size (MW)} & \textbf{Lifetime} \\
\hline
\textbf{Coal} & Coal &  &  & 236.8 &  &  & 24 & 24 & 60\% & 60\% & 55\% & 90\% &  & 30 \\
\textbf{CCGT} & Natural gas & & & 106.5 &  &  & 8 & 8 & 100\% & 100\% & 50\% & 90\% &  & 30 \\
\textbf{Nuclear} & Uranium & 0.6 & 75 & 1,000 &  & 10.1 & 36 & 36 &  &  & 90\% & 90\% & 1,000 & 40 \\
\textbf{Biomass} & Biomass & 0 & 37.88 &  &  & 16.7 & 24 & 24 & 60\% & 60\% & 55\% & 90\% & 1 & 20 \\
\textbf{New Coal} & Coal & 0.9 & 30 & 214 & 0 & 9.5 & 24 & 24 & 60\% & 60\% & 45\% & 90\% & 620 & 30 \\
\textbf{New CCGT} & Natural gas & 1.5 & 10 & 106.5 & 0 & 6.6 & 8 & 8 & 100\% & 100\% & 33\% & 90\% & 573 & 30 \\
\textbf{New OCGT} & Natural gas & 7 & 11 & 96 & 0 & 9.1 & 2 & 2 & 100\% & 100\% & 26\% & 90\% & 384 & 30 \\
\textbf{Backup} & Diesel &  & 0 & 0 & 0 & 10.9 & 0 & 0 & 100\% & 100\% &  & 90\% & &\\
\hline
\end{tabularx}
\label{table:ntl_thermal}
\end{table}

\end{landscape}

\begin{landscape}
\begin{table}[!ht]
\centering
\caption{Hydro power existing capacity and parameters \cite{NREL2020, Rudnick2019}}
\begin{tabularx}{25cm}{c|ccccc|CCcCCC}
\hline
       & \multicolumn{5}{c|}{\textbf{2020 Capacity (MW)}}      &       && \multicolumn{1}{l}{} & \multicolumn{1}{l}{}  & \multicolumn{1}{l}{}      & \multicolumn{1}{l}{}      \\
\textbf{Resource}       & \textbf{North} & \textbf{West} & \textbf{South} & \textbf{East} & \textbf{Northeast} & \textbf{VOM (\$/MWh)} & \textbf{FOM (\$/kW/y)} & \textbf{Lifetime}    & \textbf{Efficiency Up/down} & \textbf{Power to energy ratio} & \textbf{Initial hydro level (\% of reservoir)} \\
\hline
\textbf{Hydro reservoir}& 7,103   & 5,494  & 7,429   & 4,217  & 2,061       & 0     & 34.85 & 50   &      &   &   \\
North   & 7,103   &&&&&&&&& \num{8.89e-4}     & 0 \\
West    & 5,494   &&&&&&&&& \num{6.19e-4}    & 0 \\
South   & 7,429   &&&&&&&&& \num{4.18e-4}     & 0 \\
East    & 4,217   &&&&&&&&& \num{1.1e-3}     & 0 \\
Northeast       & 2,061   &       &&       &    &       &&      &       & \num{8.24e-3}     & 0 \\
\hline
\textbf{Hydro run-of-river}     & 16,235  & 1693  & 2,430   & 1,417  & 839& 0     & 34.85 & 50   &      &  &  \\
North   & 16,235  &&&&    &&&      &       &   &   \\
West    & 1,693   &&&&    &&&      &       &   &   \\
South   & 2,430   &&&&    &&&      &       &   &   \\
East    & 1,417   &&&&    &&&      &       &   &   \\
Northeast       & 839    &&&       &    &       &&      &       &   &   \\
\hline
\textbf{Pumped hydro storage} && 1,840  & 2,005   & 940   &    &       & 34.85 & 50   & 89.4\%& 0.083333  &  \\
\hline
\end{tabularx}
\label{table:hydro}
\end{table}
\end{landscape}

\begin{landscape}

\begin{table}[!ht]
\caption{Existing Variable Renewable Energy \cite{NREL2020, ATB2020, osti_1563140, DESHMUKH2019947, OSM2020, WB2020, LULC16}}
\centering
\begin{tabularx}{20cm}{ll|CCccccccc}
\hline
\multicolumn{1}{l}{}& \multicolumn{1}{l|}{} &    &     && \multicolumn{3}{c}{\textbf{Interconnection cost (\$/MW)}}  & \multicolumn{3}{c}{\textbf{Maximum capacity (MW)}} \\

\multicolumn{1}{l}{}& \multicolumn{1}{l|}{} & \multicolumn{1}{c}{\textbf{2020 capacity (MW)}} & \multicolumn{1}{p{2cm}}{\textbf{Minimum retirement in 2050 (MW)}} & \multicolumn{1}{c}{\textbf{Lifetime}} & \multicolumn{1}{c}{\textbf{Bin 1}} & \multicolumn{1}{c}{\textbf{Bin 2}} & \multicolumn{1}{c}{\textbf{Bin 3}} & \multicolumn{1}{c}{\textbf{Bin 1}} & \multicolumn{1}{c}{\textbf{Bin 2}} & \multicolumn{1}{c}{\textbf{Bin 3}} \\
\hline
\multirow{3}{*}{\rotatebox[origin=c]{90}{\textbf{Wind}}} & North& 7,267& 7,267& 25& 5,323 & 11,910& 8,357 & 1,381,894 & 572,033& 779,177\\
  & West     & 19,659    & 19,659     & 25   & 5,253 & 6,468 & 4,593 & 1,078,581 & 1,093,093 & 458,577  \\
  & South    & 22,979 & 22,979  & 25   & 6,251 & 5,645 & 5,860 & 758,149& 168,913& 745,844\\
  \hline
\multirow{5}{*}{\rotatebox[origin=c]{90}{\textbf{Solar}}}& North    & 8,393   & 8,393     & 30   & 7,724 & 24,726& 6,198 & 898,332& 114,045& 1,106,127 \\
  & West     & 10,889    & 10,889   & 30   & 5,251 & 7,997 & 5,159 & 774,465& 898,509& 402,062\\
  & South    & 21,522    & 21,522     & 30   & 5,259 & 7,041 & 5,660 & 376,212& 234,610& 265,542\\
  & East     & 1,100   & 1,102     & 30   & 4,139 & 4,656 & 4,980 & 183,627& 301,833  & 129,668\\
  & Northeast& 323  & 323   & 30   & 6,033 & 5,347 & 56,541& 152,712& 45,735& 40,998\\
  \hline
\end{tabularx}
\label{table:vre}
\end{table}
\end{landscape}

\begin{table}[!ht]
  \centering
  \caption{Zonal power transfer limits (Annex 7.1 of \cite{CEATrans2017}). Zonal definitions as shown in Supplementary Fig. \ref{fig:18_SI}.}
    \begin{tabularx}{10cm}{l|C C C}
    \hline
    \textbf{Zonal links} & \textbf{2020 zonal capacity limits (MW)} & \textbf{Distance (km)} & \textbf{Line loss} \\
    \hline
    East-North & 22,530 & 1,140 & 7.1\%\\
    West-North & 36,720 & 851 &5.3\%\\
    West-South & 23,920 & 812 &5.0\%\\
    North-Northeast & 3,000 & 1,684 &10.5\%\\
    West-East & 21,190 & 937 &5.8\%\\
    South-East & 7,830 & 1,241 &7.7\%\\
    East-Northeast & 2,860 & 863 &5.4\%\\
    \hline
    \end{tabularx}%
  \label{table:network}%
\end{table}%

%%% STEP BY STEP DESCRIPTION OF SUPPLY CURVE GENERATION - MOVE TO SI SECTION %%%%%%%%%%%%
%%%% SOME SENTENCES ARE REPEATED FROM ABOVE TO MAINTAIN CONTINUITY %%%%%
\subsection*{Variable renewable energy resource characterization}\label{method:resource-si}
Fig. \ref{fig:flowchart_SI} describes our approach to generate parameters used to parameterize VRE resource availability in the GenX model. First, we translate wind and solar resource data available for each location from the Renewable Energy Potential Model (reV)\cite{osti_1563140} into hourly normalized power output (or capacity factor (CF)) profiles. For wind, this is done using the power curve of the Siemens Gamesa 126/2500 \cite{gamesa} wind turbine with hub height 84 meters, while for solar, we model a single-axis tracking, horizontally oriented PV system using the NREL System Advisor Model\cite{SAM2020}. Second, we estimate the land area available for wind and solar generation based on 2005 land use and land coverage classifications across India \cite{LULC16} and considering constraints on elevation \cite{WB2020}. Similar to \cite{DESHMUKH2019947} we restrict land usage for solar development to: shrubland, wasteland, salt pan, grassland, while land eligible  for wind development also cropland, barren and fallow land. We further exclude parcels of land with a slope greater than 5\% for solar development and 20\% for wind development. We use $32 MW/km^2$ and $4 MW/km^2$ for spatial density of solar and wind resources respectively \cite{NREL2020} to convert available land area within each grid cell (25$ km^2$ resolution) into nameplate capacity that can be deployed. Third, we identify the closest point sampled from reV to associate a CF profile to each pixel. Fourth, we identify the cost of interconnection of each pixel by extending a straight line from the centroid of the pixel to the minimum cost substation by factoring voltage dependent cost of interconnection with distance. Here substation data is sourced from from OpenStreetMaps \cite{OSM2020}. Fifth, we  aggregate different resource sites into a small number of resource bins (3 for wind and 3 for solar for each zone) that can be represented in the GenX model based on clustering the sites using the levelized cost of energy (LCOE) metric. The LCOE for each site is computed using site-specific CF and interconnection costs as well as capital costs and Fixed O\&M costs  from NREL Annual Technology Baseline (ATB) 2020 \cite{ATB2020} (utility-scale PV and class 6 wind data). Parameter inputs developed to characterize each resource bin in the GenX model are summarized in \ref{table:vre} and include : a) hourly CF, computed as the weighted average CF for sites within each bin, where the weights correspond to the developable area associated with that site., b) total developable capacity and c) weighted average annualized interconnection cost associated with each bin. 

India's growth in the twenty-first century can most closely compared to China's with China ahead of India with respect to VRE capacity installation \cite{IEAWEO2020}. Given how China has been deploying VRE for a longer time, there are more data points to fit a Gompertz growth curve \cite{10.2307/86156}. We use the Gompertz sigmoid function to simulate slow initial adoption, rapid ramp-up followed by slow progress which is a good representation of new technology deployment. Results of the curve fitting are shown in Supplementary Table \ref{table:gompertz}. In 2019, India had 37.5 GW of wind and 33.7 GW of solar capacity installed nationally \cite{IEAWEO2020}. These capacity values are inserted in their corresponding fitted Gompertz curve to map to the year China was at that capacity. The mapped years are 2014 and 2010 for solar and wind respectively. Projecting the decadal installation limits are then extracted from the curves with the starting points being the identified mapped years. Results are shown in Supplementary Table \ref{table:caplimit}.

\begin{table}[!ht]
\centering
\caption{Gompertz curve fitting results}
\begin{tabular}{l|cccc}
\hline
  \textbf{Parameter}& \textbf{Solar} & \textbf{Wind}\\
      \hline
$A$  & 2,952.55    & -1,119.48 \\
$\mu$ & 89.53    & 39.40\\
$d$ & 2,018.93    & 2,042.84\\
$y_0$ & -5.69    & 1,042.94\\
R-squared & 0.999434 & 0.996302 \\
\hline
\end{tabular}
\label{table:gompertz}
\end{table}

\begin{table}[!ht]
\centering
\caption{Decadal VRE installation limit (MW) for the reference case}
\begin{tabular}{l|cccc}
\hline
  \textbf{Resource}& \textbf{2020} & \textbf{2030}& \textbf{2040}& \textbf{2050}\\
      \hline
Wind  & 0    & 171,000 & 320,000 & 364,000 \\
Solar & 0    & 443,000 & 854,000 & 746,000  \\
\hline
\end{tabular}
\label{table:caplimit}
\end{table}

\subsection*{Distribution-level storage modeling}\label{method:dls-si}

The distribution-level storage (DLS) demand scenario captures the transmission-level effects of mass deployment of battery storage on distribution networks as a non-wire alternative network upgrade strategy. The demand profile for each investment period under the DLS scenario is computed based on an offline assessment that combines three analytical components that define the flexible valuation framework, described in detail elsewhere \cite{BarbarRO2021}. These steps are: 1) identify location of congestion 2) A linear multi-period optimization is used to size the battery storage system given the ampere limit and hourly demand of the feeder and 3) a real options analysis using Markov Chain Monte Carlo (MCMC) method is applied to identify the time-evolution of the least-cost investment strategy between battery storage and network upgrades under demand growth uncertainty. First, we identify the optimal location to relieve the distribution network from congestion. Congestion occurs primarily during peak hours because of high simultaneity in demand for electricity, which implies that various components of the network overload simultaneously. This enables us to estimate the cost of traditional network upgrades that may be deferred. Identifying the optimal location enables to relieve the most significant number of components on the feeder from a minimum number of locations. Second, we evaluate the cost-optimal sizing of the battery storage system at the identified location for a given demand scenario, and depending on the hourly load profile as well as the thermal limits of the network components. A time-series linear program is used to size the system for the various demand growth scenarios. The result of the optimization is a battery storage system capacity from which we infer cost of DLS. This second step is repeated for the various considered demand growth scenarios (slow, stable, rapid). In the final part of the flexible valuation framework, we use an MCMC simulation of all the considered demand growths and their respective posterior probabilities to identify the expected cost-saving option value of DLS and network deferrals under demand uncertainty. MCMC simulations enable estimating parameters such as mean, variance, expected values of the posterior distribution of a Bayesian model. Given three scenarios on electricity demand growth \cite{Barbar2021}, we use MCMC Metropolis-Hastings (MH) to find the expected values of demand growth on a decadal basis \cite{BarbarRO2021}. The result of the MCMC simulation is a posterior distribution function that we use to sample the scenario probability to identify the option cost of installing DLS and deferring network investment. For the MCMC simulation, we compute the annualized investment cost based on the storage size from step 2 as well as the fixed operation and maintenance cost of the battery storage system. The real option value of flexibility will be the difference between traditional network upgrades investment and a storage system with postponed upgrades; this is the value of deferral. We use annualized investment cost for all calculations so that multiple deferrals can be considered sequentially and account for the salvage value of the battery storage when the real option value of flexibility is no longer favorable. The deferral value $D$ is calculated by summing the demand projections product of the MCMC probabilities and their object cost (i.e. storage, network upgrades as defined in Supplementary Table \ref{table:distro}). If $D < 0$, then the cost of storage and network upgrade deferral is lower than traditional network upgrades and therefore storage has NWA value. The process is repeated at every decision point $p$ with the corresponding cost, projections and, probabilities.

Supplementary Fig. \ref{fig:dls-flow} details the usage of the flexible valuation framework to estimate DLS potential at scale in select Indian megacities. Nine representative feeders are identified for analysis via clustering techniques applied to a library of urban feeders (and their respective hourly demand profiles) for the city of Delhi provided by Tata Power Delhi Distribution Limited (TPDDL) \cite{tata2019}. Each representative feeder is characterized by: a) loading percentage that varies from 40 to 80\% based on the collected data \cite{tata2019}, b) represented demand that is defined as the hourly load profile modeled on the feeder, which will vary by megacity according to available survey data \cite{prayas2012}, c) serviced demand which is the total annual demand (MWh) that is serviced by distribution network feeders with the same loading percentage and d) serviced circuit kilometers (km) which corresponds to the total circuit km that is at the corresponding loading percentage. The data from the city of Delhi shows that 28\% of the feeders were loaded at 60\% or more on an ampere capacity basis in 2018, we assume a similar distribution to the other megacities considered. Using each megacity's MWh to circuit kilometers ratio we identify the circuit kilometers for each of the nine feeders represent based on their respective serviced demand. The flexible valuation framework is applied on each representative feeder for each megacity by using the appropriate growth rates from the demand-side modeling for various demand projections in 2030 and 2040 \cite{Barbar2021}. First, network investment cost are calculated based on the circuit kilometers from each representative feeder. Second, the resulting DLS capacity for each representative feeders is scaled up to produce megacity-level estimate of DLS capacity using the feeder's serviced demand to represented demand ratio \cite{BarbarRO2021}. While the range of plausible demand projections is the widest in 2050 \cite{Barbar2021}, we focus on deferral value that DLS can provide in 2030 and 2040. The flexible valuation framework's outcomes differ based on cost (network upgrade, battery storage capital costs) and growth projections. Battery storage has limited deferral life that may be well short of its useful life, therefore assessing the flexible framework over multiple investment periods is necessary. By initially installing DLS, the decision-maker (i.e. utility) first observes demand growth realization and then commit to longer term investment. Our analysis reveals that under high uncertainty of growth projection, the flexible valuation framework increases favorability towards installing DLS when peak demand grows at a slower than anticipated pace. This allows the utility to adopt a \textit{wait and see} strategy without compromise on quality of supply. Finally, since power is mostly contracted in many cities in India, we do not consider the potential value of arbitrage that DLS can offer utilities if market conditions exist. So it is important to note that while DLS is utilized for only peak hours of the year, it can be more actively deployed and hence have a higher value than traditional network upgrades that do not have multiple use cases. 

% Dharik (05/26/2021): Note sure if this para that I copied over from the main text fits anywhere and may be needed.  IF you feel otherwise, add it at the best place you think it should go. I cut it out from the main text since it saves us some words.

%We decouple the optimizations of DLS and grid-scale storage since the DLS costs are compared to the high network upgrades cost Indian utilities are experiencing \cite{CEA18} whereas grid-scale storage costs are considered in the investment and operation of the overall system. We investigate DLS based on the trend of growing interest by Indian distribution companies. DLS is deployed as a cost saving NWA network upgrade and is not designed to inform decision on planning and coordination of storage at the transmission vs. distribution level of the electrical grid. 

Our bottom-up DLS assessment is based on deployment in four megacities that accounted for 52 TWh of annual electricity consumption in 2019 and 72,763 circuit kilometers serving dense urban areas by their respective utilities \cite{CEA18, TPDDL}. Without any distribution network expansion consideration and under stable demand projections, we estimate 20,373 km will be overloaded, i.e. over 80\% ampacity, in 2030 under the demand scenarios based on reference and an additional 23,640 km will be overloaded in 2040.  When the above flexible valuation framework is applied on the representative feeders and then scaled back to the total demand they represent, the cost-effective option is to install a total of 29 and 140 GWh of short duration storage to defer 15,914 and an additional 18,127 km of network upgrades for 2030 and 2040 respectively. As seen in Supplementary Table \ref{table:dls-results}, when considering multi-period investment deploying DLS before traditional network upgrades produces 16\% capital investment savings in 2030 and 15\% in 2040. More DLS is deployed in 2040 per unit kilometers compared to 2030 due to increasingly peaky nature of the projected demand \cite{Barbar2021}. Across the regions, DLS deployment is dominant in the Northern and Western regions in 2030 due to higher growth rates for average and peak electricity demand which leads to more congested urban feeders.Heavily loaded feeders may not accommodate battery storage, from a storage charging perspective, and therefore will require traditional network upgrades.

Our bottom-up assessment of DLS deployment is based on DLS cost projection that are in line with transmission level storage cost assumptions as highlighted in Supplementary Table \ref{table:distro} \cite{ATB2020, MOCI2019}. DLS optimization is constrained by two factors: dispatch and cost. Under the low cost of storage scenario the flexible valuation framework yields the same results for DLS which indicates that the only binding constraint is dispatch i.e. the availability of off-peak line thermal capacity on the feeder to charge DLS for peak hours discharge. It is only when costs are projected to be higher that we note a decrease in DLS installation (see \cite{BarbarRO2021} for further details). Therefore, we use the mid-range cost projections results for the supply-side modeling (Table \ref{table:dls-sens}). DLS is assumed to remain present on the system as long as it is dispatchable since the longer it remains on the feeder the more value it defers. DLS useful life as a DLS range between 5 and 10 years, with the ability to carry over from one modeling period to the next.Given that DLS is modeled as a network upgrade strategy for capital expenditure minimization at distribution-level to reduce system peak by charging from the transmission during off-peak hours, it is therefore considered a zero-cost peak shifting mechanism at the regional level.

Table \ref{table:dls} columns 1 and 2 compare the 2030 and 2040 SCOE results for the reference and DLS scenario cases, respectively. The 3$^{rd}$ column presents the total system cost of electricity for DLS only using Eqn. \ref{eqn:1} but without any transmission expansion and considering only DLS storage as a technology and the peak demand that is impacted by DLS. This figures enables an aggregate comparison of DLS cost at transmission level to assess the impact on the bulk power system.

%  dharik (05/26/2021): Explain in 2-3 sentences the findings in Table 9 in the SI on SCOE and COE

% Dharik (05/26/2021) - I dont think this needed - we can be concise here.
%As of 2018, India installed over 10 million kilometers of electric distribution lines \cite{CEA2018} with 8\% covering dense urban areas \cite{WEC2009}. Major metropolitan areas in India witness a high rate of urbanization that is stressing the distribution grids technically and the companies that operate them financially \cite{WB2014}. An increasingly simultaneous evening peak causes an overload on the feeder that may cause stability problems upstream, distribution companies must either shed load or invest in network upgrades. Shedding load is increasingly penalized and distribution network upgrades are very costly, especially reconductoring cables in densely populated areas \cite{NRELCD2019}. 

\begin{figure}[!ht]
\centering
 \includegraphics[width=15cm,height=15cm, keepaspectratio]{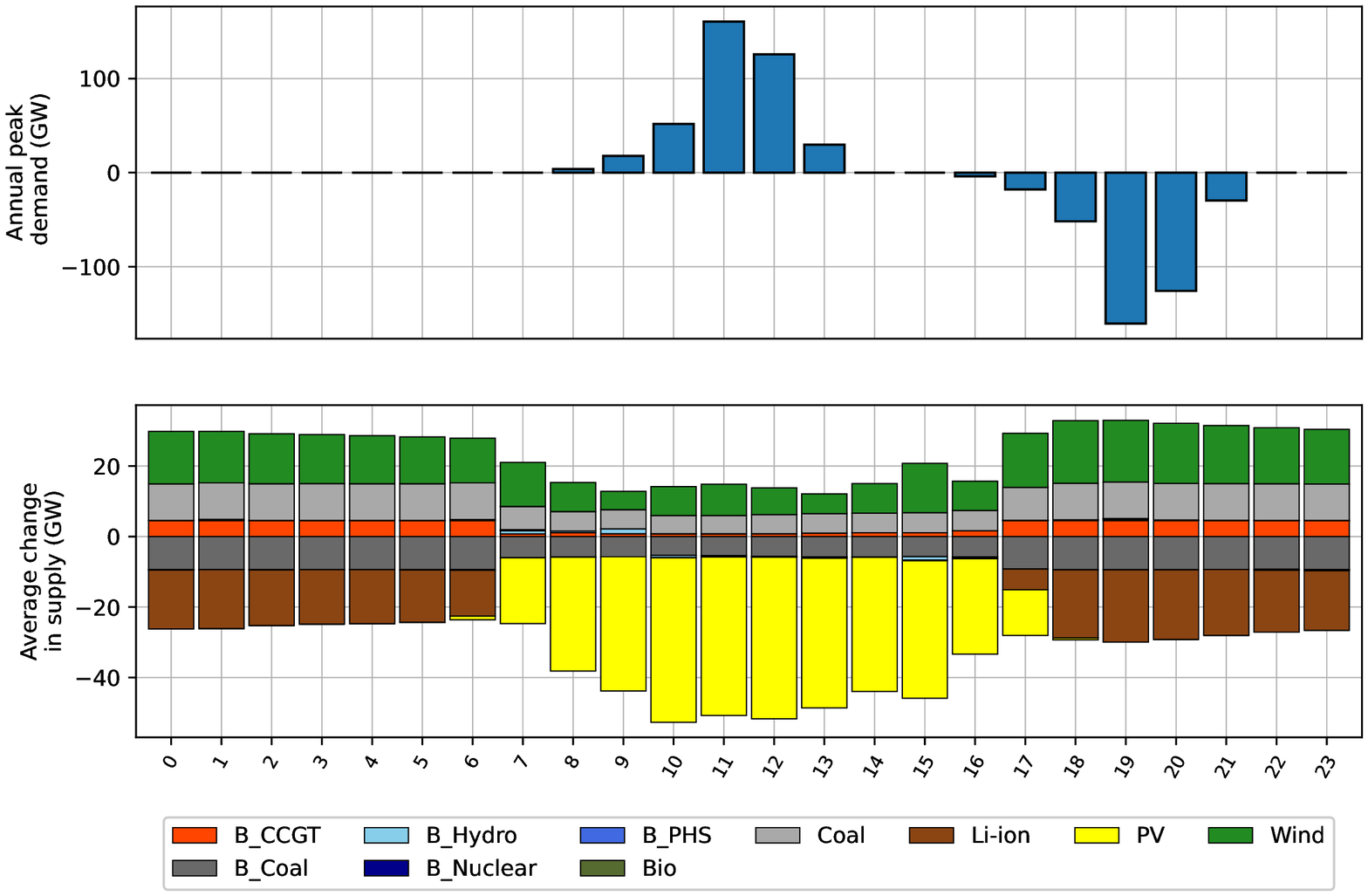}
  \caption{Yearly average DLS impact on supply and demand per hour for 2030}
   \label{fig:dls-impact}
\end{figure}

\begin{table}[!ht]
\centering
\caption{DLS system cost of electricity comparison}
\begin{tabularx}{10cm}{C|CC|C}
\hline
\multicolumn{4}{c}{\textbf{Reference case}} \\
\hline
 & Reference case & DLS case & DLS SCOE \\
 \hline
2030 & 25.9  & 25.8 & 0.4\\
2040 & 29.5 & 19.5 & 0.5 \\
\hline
\hline
\multicolumn{4}{c}{\textbf{Low cost storage case}} \\
\hline
 & DLS SCOE & DLS COE & Low storage SCOE \\
 \hline
2030 & 26.5 & 0.4 & 26.4 \\
2040 & 17.7 & 0.5 & 18\\
\hline
\end{tabularx}
\label{table:dls}
\end{table}

\begin{table}[!ht]
\centering
\caption{Storage cost impact on flexible valuation framework results for year 2030. Details available elsewhere \cite{Barbar2021}}
\begin{tabular}{l|cccc}
\hline
 & Low & Mid & High & Breakeven \\
 \hline
Storage energy cost (\$/kWh) & 116 & 168 & 236 & 261 \\
Storage power cost (\$/kWh) & 101 & 146 & 205 & 227 \\
DLS energy capacity (GWh) & 29 & 29 & 18 & 0 \\
Deferred upgrades (km) & 15,914 & 15,914 & 11,752 & 0\\
\hline
\end{tabular}
\label{table:dls-sens}
\end{table}

\begin{table}[!ht]
\centering
\caption{Flexible valuation framework aggregate results. Details available here \cite{Barbar2021}}
\begin{tabularx}{12cm}{l|CC}

\hline
Cost&\textbf{2030} &\textbf{2040} \\
\hline
DLS & \$206,938,425 & \$260,828,580 \\
Annualized deferred upgrades & \$75,879,205 & \$135,747,611  \\
Annualized traditional upgrades & \$116,749,563  & \$132,985,321 \\
\hline
Total flexible budget & \$2,931,672,259 & \$5,323,455,899 \\
Total traditional budget & \$3,502,486,892  & \$6,265,935,776 \\
\end{tabularx}
\label{table:dls-results}
\end{table}

\begin{table}[!ht]
\centering
\caption{DLS cost assumptions}
\begin{tabularx}{12cm}{l|CC}
\hline
&\textbf{2030} &\textbf{2040} \\
\hline
Energy Cost (\$/kWh) & 168 & 147 \\
Power Cost (\$/kW) & 146 & 128\\
Battery O\&M cost (\$/kW-year) & 20 & 18\\
Charge cost (\$/MWh) & 55 & 55\\
\hline
New line (\$/km) & 350,000 & 350,000\\
Reconductoring (\$/km) & 650,000 & 650,000\\
\hline
\end{tabularx}
\label{table:distro}
\end{table}

\subsection*{System cost of electricity calculation}\label{method:cost-si}

The modeled equations are as follows:
\begin{itemize}
\item Investment cost (Table \ref{table:1}):
\begin{equation}
    A_{y,t} = G_{y,t} \cdot I_{y,t} + X_{y,t} \cdot IC_{y,t} + B_{y,t} \cdot IS_{y,t} + TX_{t} \cdot T_{t}
    \label{eqn:1}
\end{equation}

\begin{table}[!ht]
\centering
\caption{Abbreviations for Eqn. \ref{eqn:1}}
\begin{tabularx}{12cm}{l|CC}
\hline
 $y$ & Model period& \\
 $t$ & Technology & Table \ref{table:abr}\\
 $A$ & Total investment cost & USD\\
 $G$ & New installed capacity & MW \\
 $I$ & Investment cost & USD/MW\\
 $X$ & New installed storage power & MW\\
 $IC$ & Investment cost of storage power & USD/MW\\
 $B$ &  New installed storage capacity & MWh\\
 $IS$ &Investment cost in storage capacity &USD/MWh\\
 $TX$ &Transmission expansion capacity&GW\\
 $T$ &Transmission expansion cost&USD/GW\\
\hline
\end{tabularx}
\label{table:1}
\end{table}

\item Annualized investment cost (Table \ref{table:2}):
\begin{equation}
    AIC_y = \sum_{t}^{tech}\bigg(A_{y,t} \cdot \frac{WACC}{1-(1+WACC)^{-L_{t}}}\bigg)
    \label{eqn:2}
\end{equation}

\item Previous period technology that is deployed in the current period annualized investment cost (Table \ref{table:2}):
\begin{equation}
    AIC_{y,p} = \sum_{t}^{tech}AIC_{p,t}
    \label{eqn:3}
\end{equation}

\begin{table}[!ht]
\centering
\caption{Abbreviations for Eqn. \ref{eqn:2} and \ref{eqn:3}}
\begin{tabularx}{12cm}{l|CC}
\hline
 $y$ & Model period& \\
 $t$ & Technology & Table \ref{table:abr}\\
 $AIC$ & Annualized investment cost & USD\\
 $A$ & Total investment cost & USD\\
 $WACC$ & Weighted average cost of capital & \% \\
 $L$ & Technology life & years \\
\hline
\end{tabularx}
\label{table:2}
\end{table}

\item Yearly fixed operation and maintenance cost (Table \ref{table:4}):
\begin{equation}
    FOM_y = \sum_{t}^{tech}\big(E_{y,t} \cdot F_{y,t} + W_{y,t} \cdot FC_{y,t} + Y_{y,t} \cdot FS_{y,t}\big)
    \label{eqn:4}
\end{equation}

\begin{table}[!ht]
\centering
\caption{Abbreviations for Eqn. \ref{eqn:4}}
\begin{tabularx}{12cm}{l|CC}
\hline
 $y$ & Model period& \\
 $t$ & Technology & Table \ref{table:abr}\\
 $FOM$ & Total fixed operation and maintenance cost & USD\\
 $E$ & Available generation size & MW\\
 $F$ & Fixed operation and maintenance cost & USD/MW-year \\
 $W$ & Available storage power & MW \\
 $FC$ & Fixed operation and maintenance cost of storage power & USD/MW-year\\
 $Y$ & Available storage capacity & MWh \\
 $FS$ & Fixed operation and maintenance cost of storage capacity & USD/MWh-year\\
\hline
\end{tabularx}
\label{table:4}
\end{table}

\item Yearly variable cost (Table \ref{table:5}):
\begin{equation}
    VAR_y = \sum_{t}^{tech}\big(D_{y,t} \cdot V_{y,t} + Q_{y,t} \cdot VC_{y,t}\big)
    \label{eqn:5}
\end{equation}

\begin{table}[!ht]
\centering
\caption{Abbreviations for Eqn. \ref{eqn:5}}
\begin{tabularx}{12cm}{l|CC}
\hline
 $y$ & Model period & \\
 $t$ & Technology & Table \ref{table:abr}\\
 $VAR$ & Variable cost & USD\\
 $D$ & Annual generation & MWh\\
 $V$ & Variable operation and maintenance cost & USD/MWh \\
 $Q$ & Total storage charging capacity & MWh \\
 $VC$ & Variable operation and maintenance charge cost & USD/MWh\\
\hline
\end{tabularx}
\label{table:5}
\end{table}

\item Fuel cost (Table \ref{table:6}):
\begin{equation}
    U_y = \sum_{t}^{tech}\big(D_{y,t} \cdot P_{y,t} \cdot H_{y,t}\big)
    \label{eqn:6}
\end{equation}

\begin{table}[!ht]
\centering
\caption{Abbreviations for Eqn. \ref{eqn:6}}
\begin{tabularx}{12cm}{l|CC}
\hline
 $y$ & Model period & \\
 $t$ & Technology & Table \ref{table:abr}\\
 $U$ & Total fuel cost & USD\\
 $P$ & Fuel price & USD/MMBtu\\
 $H$ & Heat rate & MMBtu/MWh \\
\hline
\end{tabularx}
\label{table:6}
\end{table}

\item Startup cost, startup fuel cost and start cost (Table \ref{table:7}):
\begin{equation}
    S_y = \sum_{t}^{tech}\big(R_{y,t} + O_{y,t}\big)
    \label{eqn:7}
\end{equation}

\begin{equation}
    R_{y,t} = SF_{y,t} \cdot N_{y,t} \cdot P_{y,t} \cdot Z_{y,t}
    \label{eqn:8}
\end{equation}
\begin{equation}
    O_{y,t} = SC_{y,t} \cdot N_{y,t} \cdot Z_{y,t}
    \label{eqn:9}
\end{equation}

\begin{table}[!ht]
\centering
\caption{Abbreviations for Eqn. \ref{eqn:7}, \ref{eqn:8}, \ref{eqn:9}}
\begin{tabularx}{12cm}{l|CC}
\hline
 $y$ & Model period & \\
 $t$ & Technology & Table \ref{table:abr}\\
 $S$ & Total startup cost & USD\\
 $R$ & Total start fuel cost & USD\\
 $O$ & Base start cost & USD \\
 $SF$ & Start fuel & MMBtu/MW\\
 $N$ & Number of starts in a year & \\
 $Z$ & Generation capacity size & MW\\
 $SC$ & Base start cost & USD/MW \\
\hline
\end{tabularx}
\label{table:7}
\end{table}

\item System cost of electricity (Table \ref{table:10}):
\begin{equation}
    SCOE_y = \frac{AIC_y+\sum_{p \in M} AIC_{y,p}+FOM_y+VAR_y+U_y+S_y}{D_y}
    \label{eqn:10}
\end{equation}

\begin{table}[!ht]
\centering
\caption{Abbreviations for Eqn. \ref{eqn:10}}
\begin{tabularx}{12cm}{l|CC}
\hline
 $y$ & Model period & \\
 $t$ & Technology & Table \ref{table:abr}\\
 $p$ & Previous model period & \\
 $M$ & Set of model periods & Table \ref{table:capcost} \\
 $SCOE$ & System cost of electricity & USD/MWh\\
 $AIC$ & Annualized investment cost & USD\\
 $FOM$ & Total fixed operation and maintenance cost & USD\\
 $VAR$ & Variable cost & USD\\
 $U$ & Total fuel cost & USD\\
 $S$ & Total startup cost & USD\\
 $D$ & Total electricity demand & MWh\\

\hline
\end{tabularx}
\label{table:10}
\end{table}

\end{itemize}

\bibliography{sample}